\documentclass[a4paper,11pt]{article}
\pdfoutput=1
\usepackage{jheppub}
\usepackage{array}  

\usepackage[T1]{fontenc} 

\usepackage{amssymb, amsmath, amsthm,latexsym}
\usepackage[english]{babel} 
\usepackage[latin1]{inputenc} 
\usepackage{graphicx}
\usepackage[all]{xy}
\usepackage{bbm}
\usepackage{psfrag}

\newcommand{\be}{\begin{equation}}
\newcommand{\ee}{\end{equation}}
\def\bea#1\eea{\begin{align}#1\end{align}}
\newcommand{\nn}{\nonumber}

\newcommand{\w}{\wedge}

\def\d {{\rm d}}
\def\a {\alpha}
\def\b {\beta}
\def\g {\gamma}

\newcommand\ie{i.e.}
\newcommand\eg{e.g.}

\newcommand{\cW}{\mathcal W}

\newcommand{\cD}{\mathcal D}

\newcommand{\cN}{\mathcal N}

\newcommand{\cM}{{\cal M}}

\title{\boldmath  Revisiting toric $SU(3)$ structures}
\author[]{Magdalena Larfors}

\affiliation[]{Mathematical Institute, University of Oxford \\
Andrew Wiles Building, Radcliffe Observatory Quarter\\ 
Woodstock Road, Oxford, OX2 6GG, England}

\emailAdd{larfors@maths.ox.ac.uk}
\abstract{Three-dimensional smooth compact toric varieties (SCTV) admit $SU(3)$ structures, and may thus be relevant for string compactifications, if they have even first Chern class ($c_1$). This condition can be fulfilled by infinitely many SCTVs, including $\mathbb{CP}^3$ and $\mathbb{CP}^1$ bundles over all two-dimensional SCTVs. We show that as long as $c_1$ is even, toric $SU(3)$ structures can be constructed using a method proposed in \cite{Larfors:2010wb}. We perform a systematic study of the parametric freedom of the resulting $SU(3)$ structures, with a particular focus on the metric and the torsion classes. Although metric positivity constrains the $SU(3)$ parameters, we find that every SCTV admits several toric $SU(3)$ structures and that parametric choices can sometimes be made to match requirements of string vacua. We also provide a short review on the constraints that an $SU(3)$ structure must meet to be relevant for four-dimensional, maximally symmetric $\cN=1$ or $\cN=0$ string vacua.}
\arxivnumber{1309.2953}
\keywords{Flux compactification, Toric variety}
\begin{document}

\maketitle

\flushbottom

\section{Introduction}
\label{sec:intro}

As is well-known, string and M theory are higher dimensional theories, that can yield phenomenologically relevant four-dimensional models through compactification. The vast majority of these constructions use a Calabi--Yau (CY) manifold as the internal, compact space. There are many reasons for this: historically, the first four-dimensional, maximally symmetric and supersymmetric vacua were found by compactification of heterotic string theory on CY threefolds \cite{Candelas:1985en}, and since then the collective efforts of mathematicians and physicists have led to a deep understanding of these spaces. However, CY manifolds are not the most general manifolds that lead to supersymmetric four-dimensional vacua through string compactifications. Furthermore, once background fluxes and sources are introduced in the construction, CY manifolds generically fail to solve the Killing spinor equations required for supersymmetric vacua.\footnote{Calabi--Yau compactifications inevitably have moduli, which lead to phenomenological problems in the associated four-dimensional theory. Background fluxes provide one way of stabilising these moduli. For recent reviews on flux compactifications, see \cite{Grana:2005jc,Douglas:2006es,Blumenhagen:2006ci,Koerber:2010bx}.} Thus, by only focusing on CY manifolds, phenomenologically interesting flux compactifications might be missed, and premature conclusions on the properties of generic string vacua will be drawn.

$SU(3)$ structure manifolds provide a natural generalisation of CY manifolds; both types of manifolds allow a globally defined spinor, that reduces their structure groups to $SU(3)$. On CY manifolds, the spinor is in addition covariantly constant (with respect to the Levi--Civita connection), thus reducing the holonomy to $SU(3)$. A well-defined spinor is certainly needed to construct supersymmetric four-dimensional vacua, but demanding that it is covariantly constant is not necessary. It can be shown that the covariant derivative of the spinor vanishes if and only if the intrinsic torsion of the $SU(3)$ structure manifold is zero. This statement can be reformulated in terms of differential forms: bilinears of the spinor define a real two-form $J$ and a complex decomposable (3,0)-form $\Omega$ that fulfil
\be
\Omega\wedge J=0 \; , \;  
\Omega\wedge\overline{\Omega}=-\frac{4i}{3}J^3\neq 0 \; ,
\ee
and are closed if and only if the torsion vanishes. Loosely speaking, the intrinsic torsion, which can be decomposed into five torsion classes $\cW_i$, thus measures how far the manifold is from being CY.  The precise definition of the torsion classes can be found in section \ref{sec:su3constr}.
 
The intrinsic torsion means that the properties of generic $SU(3)$ structure and CY manifolds differ radically. The first two torsion classes, $\cW_1$ and $\cW_2$, are associated with the Nijenhuis tensor of the manifold, and as long as either of them is non-zero the almost complex structure of the manifold fails to be integrable. This implies that generic $SU(3)$ structure manifolds cannot be analysed using algebraic geometry. As a consequence, it has proven quite difficult to construct explicit examples of $SU(3)$ structure manifolds, and this scarcity of examples has left important aspects of flux compactifications in obscurity. While four-dimensional vacua can be found, the effective field theories that describe fluctuations around these vacua are difficult to obtain. Indeed, some properties of the vacua, such as the existence of moduli, are often hard to determine. To shed more light on these constructions, a better understanding of the dimensional reduction on $SU(3)$ structure manifolds is required. Finding more examples is of essence to meet this goal.
 
As the non-integrability of the almost complex structure is an obstacle in the construction of examples, much would be gained if already known complex manifolds could be shown to admit a second, non-integrable almost complex structure, that is associated with an $SU(3)$ structure. This has indeed been demonstrated for twistor spaces and was used by Tomasiello to show that $\mathbb{CP}^3$ and $\mathbb{CP}^1 \hookrightarrow \mathbb{CP}^2$ allow a half-flat $SU(3)$ structure \cite{Tomasiello:2007eq} (see also \cite{Xu:2006math}). Since both $\mathbb{CP}^3$ and $\mathbb{CP}^1 \hookrightarrow \mathbb{CP}^2$ are smooth, compact toric varieties (SCTV), it was proposed by L\"ust, Tsimpis and the present author that other toric varieties may also admit two almost complex structures, and that $SU(3)$ structures may exist also on these varieties \cite{Larfors:2010wb}. Since there are infinitely many SCTVs, such a construction holds the promise of substantially expanding the set of $SU(3)$ structure examples.
 
The purpose of this paper is to extend the studies of \cite{Larfors:2010wb} in two respects. First, we discuss the topological constraint that a manifold must fulfil to allow an $SU(3)$ structure. In order to allow a nowhere vanishing spinor, the second Stiefel--Whitney class of the manifold must be trivial. This can be reformulated as a constraint on the first Chern class $c_1$: only manifolds with even $c_1$ allow $SU(3)$ structures. On a toric variety, $c_1$ is easily computed as a sum of divisors, rendering the check of this topological constraint almost trivial, as recently noted by Dabholkar \cite{Dabholkar:2013qz}. We will extend the analysis of \cite{Dabholkar:2013qz} to different classes of SCTVs studied by Oda \cite{oda}, and show that $SU(3)$ structures can be constructed on all toric $\mathbb{CP}^1$ fibrations. This is an infinite number of varieties.

Our second objective is to study the torsion classes of toric $SU(3)$ structures. We construct the defining forms $J$ and $\Omega$ following the method of \cite{Larfors:2010wb}. In order to decompose $\d J$ and $\d \Omega$ into torsion classes we need to compute contractions, which requires a manageable expression for the $SU(3)$ structure metric. This metric is in general different from the metric that the SCTV inherits from the ambient space $\mathbb{C}^n$. We will provide a compact expression for it in section \ref{sec:metric}, that will be used to explicitly compute all torsion classes for example varieties in section \ref{sec:Kuniq}. 

In any string theory compactification, the torsion classes will be constrained by the supersymmetry variations, equations of motion and Bianchi identities. In particular, there is an interesting and useful connection between the supersymmetry variations and the geometry: imposing that the variations are zero leads to necessary conditions on the $SU(3)$ structure. A similar reasoning can be made for certain non-supersymmetric vacua. Consequently, once we have computed the torsion classes, we can see if the manifold is suitable for knows string compactifications. In addition to investigating if $SU(3)$ structures meet such necessary constraints for example SCTVs, we will check whether there are choices for the parameters of the construction that imply that such conditions are met in general. We find that the metric play an important role in this analysis, as additional parameter bounds arise from the requirement of metric positivity. 

The rest of this paper is organised as follows. Section \ref{sec:survey}  contains a brief survey on the role of $SU(3)$ structure manifolds in the construction of string vacua, that is provided for readers less familiar with this literature. In section \ref{sec:sctv} we present the smooth compact toric varieties, and in section \ref{sec:toricsu3} we discuss topological conditions for, and review the construction of, $SU(3)$ structures on these manifolds.  The almost complex structure and metric of the $SU(3)$ structure are computed in section \ref{sec:metric}, and general properties of the torsion classes on SCTVs are discussed in section \ref{sec:torsion}. $SU(3)$ structures on specific example manifolds are discussed in section \ref{sec:Kuniq}, and the explicit torsion classes are presented for two examples. Section \ref{sec:conclusion} contains concluding remarks and ideas for future work. Appendix \ref{sec:Mori} reviews some aspects of K\"ahler and Mori cones that are used in the construction of $SU(3)$ structures. Our conventions regarding differential forms, wedge products and contractions follow \cite{Gray:2012md}, and are summarised in appendix A of that paper.

\section{Which $SU(3)$ structures are relevant for string vacua?}
\label{sec:survey}

Since the seminal work of Strominger \cite{Strominger:1986uh}, manifolds with torsionful $SU(3)$ structures have been used to construct string vacua. The literature on the subject is by now quite vast, and it can be difficult to keep track on the constraints that are relevant for different compactifications. To put our exploration of toric $SU(3)$ structures in context, we therefor recall some of these constructions. In these scenarios, the ten-dimensional Killing spinor equations and/or equations of motion, in combination with the Bianchi identities, lead to torsion constraints that can be effectively derived using the language of generalised geometry \cite{Hitchin:2000jd,h02,g04}. We review these constraints here. To keep the length of this section within reasonable boundaries, we limit our survey to maximally symmetric four-dimensional vacua, and do not discuss non-classical corrections to the solutions.\footnote{This brief review cannot make justice to all the work done on flux compactifications on $SU(3)$ structure manifolds, and  leaves out constructions using $SU(2)$ and $SU(3)\times SU(3)$ structures. A more complete list of references can be found in \cite{Grana:2005jc,Douglas:2006es,Blumenhagen:2006ci,Koerber:2010bx}. $SU(3)$ structure manifolds also play a key role in heterotic domain wall compactifications, where the constraints on the torsion classes were recently derived in \cite{Lukas:2010mf,Gray:2012md}.} 

\subsection{Maximally symmetric $\cN=1$ vacua}

\begin{table}[tb]
\begin{tabular}{| l | l | l |}
\hline
String vacuum & Vanishing torsion classes & $SU(3)$ type  \\
\hline
\hline
 Heterotic, Type IIB ($\cN=1$ Mkw) & $\mathcal{W}_1, \mathcal{W}_2$& Complex\\
\hline
Type IIA ($\cN=1$ Mkw) & $\mathcal{W}_{1}, \mathcal{W}_3, \mathcal{W}_4$ & Symplectic \\
\hline
 Type IIA ($\cN=1$ AdS) & $\mathcal{W}_{3}, \mathcal{W}_4, \mathcal{W}_5$ & Restricted half-flat \\
\hline
\end{tabular}
\caption{\it The Killing spinor equations for four-dimensional $\cN=1$ string vacua require that the $SU(3)$ torsion classes satisfy the constraints listed in this table.  These torsion class constraints are necessary but not sufficient.} \label{tab:su3N1}
\end{table}

The $SU(3)$ structure manifolds that lead to maximally symmetric $\cN=1$ vacua can be completely classified. The $\cN=1$ Killing spinor equations are very constraining, and give necessary conditions for the torsion classes that are summarised in table \ref{tab:su3N1}. In addition, integrability statements can be made that show that the equations of motion are implied by the supersymmetry constraints and Bianchi identities \cite{Lust:2004ig,Gauntlett:2005ww,Koerber:2007hd}. Thus, it is enough to solve the latter to show that a string vacuum exists. Since supersymmetry guarantees stability, such vacua are non-tachyonic but may have flat directions.
 
The L\"ust and Tsimpis vacua of type IIA supergravity \cite{Lust:2004ig} are arguably the simplest string vacua on $SU(3)$ structure manifolds (see also \cite{Behrndt:2004km,Behrndt:2004mj}). These are four-dimensional, AdS vacua that preserve $\cN=1$ supersymmetry.\footnote{Here ``four-dimensional'' primarily indicates that the metric is block-diagonal, as these solutions, and the $\cN=0$ AdS vacua discussed in the next section, generically lack a separation of scales \cite{Tsimpis:2012tu,McOrist:2012yc}. } If the fluxes of type IIA supergravity are chosen accordingly, it can be shown that the Killing spinor equations are solved by $SU(3)$ structures that are restricted half-flat, i.~e.~$\cW_3, \cW_4$ and $\cW_5$ are all zero. The Bianchi identities for the background fluxes further impose a differential constraint on $\cW_2$. Using the integrability results discussed above, it can be shown that the constraints on the torsion classes are necessary and sufficient. 

It is also possible to construct $\cN=1$ Minkowski vacua on $SU(3)$ structure manifolds. The oldest vacua of this type are the Strominger solutions of heterotic string theory \cite{Strominger:1986uh}, which require $SU(3)$ structure manifolds that are complex, $\cW_1 = \cW_2 = 0$, and have exact $\cW_4 = 2 \cW_5$ \cite{Cardoso:2002hd}. If the third torsion class is non-zero, there must be a non-zero Neveu--Schwarz (NSNS) flux $H$, whose Bianchi identity yields an extra constraint on the geometry. 

The first $\cN=1$ Minkowski vacua of type II string theory compacitfied on $SU(3)$ structure manifolds was found by Giddings, Kachru and Polchinshi (GKP) \cite{Giddings:2001yu}, and the full set of such vacua has been classified by Gra\~na and collaborators \cite{Grana:2004bg}. In addition to solutions of the Strominger type for both type II theories, type IIA allows $\cN=1$ Minkowski vacua on symplectic manifolds, $\cW_1 = \cW_3 = \cW_4 = 0$, if furthermore $\cW_5$ is exact. To circumvent the Maldacena--Nunez no-go theorem  \cite{Maldacena:2000mw}, orientifold six-planes (O6) have to be added to the construction. Type IIB allows $\cN=1$ Minkowski vacua on complex manifolds, $\cW_1 = \cW_2 = 0$. With O3 planes, $\cW_3$ must also be zero and $\cW_4$ and $\cW_5$ are proportional (this includes the GKP vacua). If instead O5 planes are used, $\cW_3$ need not be zero.   For all three type II $\cN=1$ vacua , the Bianchi identities give differential constraints on the torsion classes,  in addition to the necessary constraints just discussed.

Before we close this section, a few comments on orientifolds are in order. As already mentioned, there are no-go theorems in flux compactifications; a four-dimensional space time with non-negative cosmological constant is only possible when sources balance the charge and tension induced by the flux in the compact space
 \cite{Maldacena:2000mw}. In type II string theory, O$p$ planes provide such sources, and are thus necessary ingredients in the Minkowski vacua just described. Moreover, approximating the O$p$ planes as smeared sources relaxes the differential conditions on the torsion classes that come from the Bianchi identities, so that solutions are easier to find. This approximation has been used to construct examples of both the  supersymmetric solutions just discussed, and the non-supersymmetric ones we turn to in the next section. Whether such smeared orientifolds can be localised is, however, not always clear, and recent discussions of this issue can be found in \cite{Blaback:2010sj,Saracco:2012wc,McOrist:2012yc,Maxfield:2013wka}.

\subsection{Maximally symmetric $\cN=0$ vacua}
A complete classification of the $SU(3)$ structures that are relevant for non-supersymmetric string vacua does not exist. These vacua are more difficult to analyse than their supersymmetric cousins; to find generic solutions one must solve the second-order ten-dimensional equations of motion, rather than the first-order Killing spinor equations. In addition to this increased complexity, there is no guarantee for the stability of generic vacua.

\begin{table}[tb]
\hspace{-0.5cm}
\begin{tabular}{| l | l | l |}
\hline
String vacuum & Constraints on torsion classes & $SU(3)$ type  \\
\hline
\hline
Type IIB (O3) &$\cW_1, \cW_2, \cW_3$ vanishing; $3\cW_4=2\cW_5$ exact & Conformal CY\\
\hline
Type IIB (O5) &$\cW_2 = 2 \cW_1(J_B-2J_F)$; $\cW_4=0$; $\cW_5$ exact & 2/4 split: $J = J_B+J_F$\\
\hline
Type IIA (O6) &$\cW_3 = \frac{3}{2} \cW_1(\mbox{Im}\Omega-4\mbox{Im}\Omega_B)$; $\cW_4=0$; $\cW_5$ exact& 3/3 split: $\Omega = \Omega_B+\Omega_F$\\
\hline
Heterotic  & $\cW_2=\cW_1(J_B-2J_F)$;  $\cW_5=2\cW_4$ exact & 2/4 split: $J = J_B+J_F$\\
\hline
\end{tabular}
\caption{\it Four-dimensional $\cN=0$ Minkowski string vacua of no-scale type require calibrated $SU(3)$ structures of the type listed in this table. Some calibrations give the manifold a fibration structure that splits $J$ and $\Omega$ into components along the base and fibre. These torsion class constraints are necessary but not sufficient.} 
\label{tab:su3N0}
\end{table}

On $SU(3)$ structure manifolds, however, one can construct classes of maximally symmetric $\cN=0$ vacua that break supersymmetry in a controllable way. By only giving up a subset of the Killing spinor equations, one can obtain Minkowski vacua whose stability is guaranteed by a no-scale structure (\ie~the potential is positive semidefinite; relaxing this leads to weaker torsion class constraints). More specifically, these $\cN=0$ vacua admit stable D or NS5 branes, a condition that can be rephrased in terms of calibrations. GKP constructed the first vacuum of this type by compactifying type IIB/F-theory with O3 planes on conformally CY manifolds, \ie~an $SU(3)$ structure manifold with $\cW_1, \cW_2, \cW_3$ vanishing and  $3\cW_4=2\cW_5$ exact \cite{Giddings:2001yu}. Additional type II vacua of this type were studied by Camara and Gra\~na \cite{Camara:2007cz}, and classified using calibrations by L\"ust and collaborators \cite{Lust:2008zd}. Similar solutions have been found in heterotic string theory \cite{Held:2010az}, and we summarise the calibration conditions for the torsion classes in table \ref{tab:su3N0}. For these non-supersymmetric vacua, the integrability results are weakened and not all equations of motion are implied by the Killing spinor equations and Bianchi identities. Consequently, both the Bianchi identities and one constraint from the equations of motion must be checked, in addition to the conditions in table \ref{tab:su3N0}. Once these constraints are satisfied, the stability of the vacua is guaranteed. 

Calibrated $\cN=0$ AdS vacua can also be found on $SU(3)$ structure manifolds. Romans constructed AdS vacua of massive type IIA supergravity using complex ($\cW_1=0=\cW_2$) or nearly K\"ahler (only $\cW_1$ non-vanishing) $SU(3)$ structures \cite{Romans:1985tz}. An extensive study of source-free type IIA vacua on nearly K\"ahler manifolds can be found in \cite{Lust:2008zd}. 
  
Finally, $\cN=0$ maximally symmetric solutions can be of dS type. While being phenomenologically very interesting, these vacua are extremely difficult to control: they are necessarily non-supersymmetric and there is no guarantee for their perturbative stability.\footnote{Recall that dS vacua are at most metastable in theories that also allow Minkowski and AdS vacua.} Thus, for every putative vacuum, one must check if it has tachyonic directions. This analysis is model-dependent and four-dimensional, and does not result in torsion class constraints. Nevertheless, by focusing on moduli that are common for sets of models, generic no-go theorems can be derived  \cite{Ihl:2007ah,hktt08,Caviezel:2008tf,Flauger:2008ad,Haque:2008jz,Shiu:2011zt}. For type IIA compactifications with O6 planes, it has been argued that Neveu--Schwarz and Ramond--Ramond fluxes (including a Romans mass) and a negative scalar curvature for the internal manifold are needed to avoid the no-go theorems \cite{Haque:2008jz}.\footnote{See \cite{Silverstein:2007ac} for an early discussion on how a negative scalar curvature helps in achieving four-dimensional dS solutions, and \cite{Douglas:2010rt} for a general discussion on compactifications on negatively curved manifolds.}  On an $SU(3)$ structure manifold, the curvature is given by the torsion classes \cite{Bedulli2007}
\be
2\mathcal{R} = 15 |\cW_1|^2 - |\cW_2|^2 - |\cW_3|^2 +
8\langle \cW_5, \cW_4 \rangle - 2|\cW_4|^2 + 4 \d * (\cW_4 + \cW_5) \; .
\ee
$\mathcal{R}$ can be positive or negative: the nearly-K\"ahler case is an example of the first, and the symplectic case is often an instance of the latter. 

Despite these caveats, proposals exist for type IIA dS vacua on $SU(3)$ structure manifolds.  In a study by Andriot and collaborators, dS solutions were found on a symplectic solvmanifold with vanishing $\cW_5$,\footnote{The stability of the solution is not demonstrated.} and it was noted that dS solutions might also be allowed on less constrained manifolds, which only require constant $\cW_1$ and vanishing $\cW_4$  \cite{Andriot:2010ju}. Another approach has been taken by Danielsson and collaborators, who have analysed dS solutions on manifolds with half-flat $SU(3)$ structures \cite{Danielsson:2009ff,Danielsson:2011au}. Examples of such solutions have been found on half-flat coset manifolds with vanishing $\cW_2$, but all suffer from perturbative instabilities \cite{Caviezel:2008tf,Danielsson:2011au}.\footnote{Complementary four-dimensional studies demonstrate that these dS solutions are not included among the four-dimensional gauged supergravities that have stable dS vacua \cite{Danielsson:2012by}.}

\section{Smooth compact toric varieties} \label{sec:sctv}

In this section we summarise the construction of smooth compact toric varieties, largely following  \cite{Larfors:2010wb} to which we refer for more details. Toric varieties are usually discussed in terms of fans (or polytopes).\footnote{A standard reference on toric geometry is \cite{fulton}, and for recent physicist-friendly reviews we refer the reader to \cite{Reffert:2007im,Denef:2008wq,Knapp:2011ip}, and section 2 in \cite{Chialva:2007sv}.} Alternatively, they can be described as the supersymmetric moduli space of a gauged linear sigma model (GLSM). If $\{z^i, ~i=1,\dots n\}$ are holomorphic coordinates of $\mathbb{C}^n$, let
\be \label{u1action}
z^i\longrightarrow e^{i\varphi_aQ^a_i}z^i 
\ee
be a $U(1)^s$ action on $\mathbb{C}^n$. The symplectic quotient 
\be \label{kmoddef}
\mathcal{M}_{2d}=\{
z^i\in\mathbb{C}^n | \sum_{i=1}^nQ^a_i|z^i|^2=\xi^a
\}/U(1)^s~
\ee
then defines a toric variety of real dimension $2d = 2(n-s)$. This is a complex variety, with local holomorphic coordinates given by $U(1)$ invariant combinations of $z^i$. The $U(1)$ charges $Q^a_i$, which completely determine the toric variety, are related to the fundamental generators $v_i$ of the fan associated to the toric variety by
\be \label{connection}
\sum_{i=1}^nQ_i^{a} v_i=0~,
\ee
for $a=1,\dots,s=n-d$. Using this relation, one can pass between the GLSM and fan descriptions of an SCTV. The fan description is useful when classifying toric varieties, as it translates features like smoothness and compactness into easily accessible properties of the fan. 

The GLSM description, on the other hand, facilitates the discussion of SCTV $SU(3)$ structures since differential forms can easily be constructed. Any differential form $\Phi$ on $\mathbb{C}^n$ restricts to a well-defined form $\Phi|$ on  $\mathcal{M}_{2d}$  if it is vertical
\be \label{vertical}
\iota_{V^a}\Phi=\iota_{\bar{V}^a}\Phi=0 \; ,
\ee
for $a=1,\dots, s$, and invariant
\be \label{invariant}
\mathcal{L}_{\mathrm{Im}V^a}\Phi=0 \; .
\ee
Here $\mathcal{L}_{V}$ is the Lie derivative with respect to the 
vector $V$, and $V^a$ are the holomorphic vector fields that generate the $U(1)^s$ action
\be \label{holvec}
V^a:=\sum_{i}Q^a_iz^i\partial_{z_i} \; .
\ee

In particular, the toric variety inherits a K\"ahler form from the standard K\"ahler form of $\mathbb{C}^n$, by projecting to its vertical component 
\be \label{eq:Jtilde}
\widetilde{J}:= \frac{i}{2} P\left(\sum_{i=1}^n \d z^i \wedge \d \bar{z}^i \right) = \frac{i}{2} \sum_{i=1}^n \cD z^i \wedge \cD \bar{z}^i \; ,
\ee
where $P$ is a projector, and the vertical component of $\d z^i$ is denoted by $\mathcal{D}z^i$. Explicitly, we have 
\be \label{Dz}
\mathcal{D}z^i=P_{ij}dz^j \; ,
\ee
where
\be \label{projector}
P_{ij}
=\delta_{ij}-Q^a_iQ^b_j\tilde{g}_{ab} z^i\bar{z}^j ~~~~~
\mbox{ (no sum on $i, j$)}
\ee
and $\tilde{g}_{ab}$ is the inverse of the real symmetric matrix
\be \label{gab}
g_{ab}=\sum_i Q^a_iQ^b_i|z^i|^2 \; .
\ee
Although $\mathcal{D}z^i$ are not globally defined on the toric variety, it is straightforward to check that the combination $\widetilde{J}$ is both vertical and invariant. Furthermore, the restriction $\widetilde{J}|$ of $\widetilde{J}$  is closed, since any well-defined form satisfies
\be \label{arm}
d(\Phi|)=P(d\Phi)|
~.
\ee
A set of well-defined one-forms is given by $\bar{z}_i\mathcal{D}z^i$. Naturally, there can only be three linearly independent (1,0)-forms on a manifold of three complex dimensions, a fact that is ensured by the moment maps which lead to the constraints
\be
\sum_{i=1}^nQ^a_i\bar{z}_i\mathcal{D}z^i = 0~;~~~a=1,\dots,d-n~.
\ee
These constraints are imposed when restricting forms to the SCTV.

In summary, a smooth compact toric variety $\cM$ is complex and K\"ahler, with K\"ahler form $\widetilde{J}|$ and metric $\tilde{G}$ inherited from the corresponding canonical structures on the ambient space. The Fayet--Iliopoulos parameters $\xi^a$ in \eqref{kmoddef} are the K\"ahler moduli of the variety, and so \eqref{kmoddef} really describes a family of toric varieties. Inside the K\"ahler cone in the moduli space, $\xi^a$ are larger than zero and the manifold is regular.  Moreover, the Betti numbers of any $d$-dimensional toric variety are known: the odd ones are all zero and the even ones are given by 
\be
b_{2k} = \sum_{j=k}^d (-1)^{j-k} {j \choose k} d_{d-j} \; ,
\ee
where $d_k$ is the number of $k$-dimensional cones in the fan (see section 4.5 in \cite{fulton} for a proof).

The triplet $(\cM, \widetilde{g}, \widetilde{J}|)$ defines a $U(3)$ structure on $\cM$. In the next section we will discuss when the structure group can be further reduced to $SU(3)$, but before we do so, we recall that three-dimensional smooth compact toric varieties with up to eight fundamental generators have been classified by Oda \cite{oda}. This classification is based on the  weighted triangulations that is created as a two-sphere intersects the three-dimensional fan associated to an SCTV, and was reviewed in detail in \cite{Larfors:2010wb}. Here we focus on three types of three-dimensional SCTVs that are specified by their $U(1)$ charges 
\begin{itemize}
\item $\mathbb{CP}^3$\\
\be \label{eq:qcp3}
Q = \begin{pmatrix}1&1&1&1\end{pmatrix}\; ,
\ee
\item $\mathbb{CP}^2$ bundles over $\mathbb{CP}^1$\\
\be \label{eq:qab}
Q=\begin{pmatrix}
1&1&a&b&0\\
0&0&1&1&1
\end{pmatrix}\; ,
\ee
where $a$, $b$ are integers specifying the 
`twisting' of the $\mathbb{CP}^2$ bundle.
\item $\mathbb{CP}^1$ bundles over two-dimensional SCTVs\\
\be \label{eq:qci}
Q=\begin{pmatrix} 
q^1_1&\dots&q^1_{n-2}&n^1&0\\
\dots & \dots & \dots & \dots& \dots\\
q^{s-1}_1&\dots&q^{s-1}_{n-2}&n^{s-1}&0\\
0&\dots&0&1&1 
\end{pmatrix}\; ,
\ee
where $n^a\in\mathbb{Z}$, $a=1,\dots,s-1$, are integers specifying the `twisting' of the $\mathbb{CP}^1$ bundle. The charge components $q^a_i$ are the $U(1)$ charges of the two-dimensional SCTV.
\end{itemize}
Note that the last class of SCTVs is infinite, as there are infinitely many two-dimensional SCTVs. These two-dimensional varieties are completely classified, and can be constructed by blow-ups of $\mathbb{CP}^2$ or the Hirzebruch surface $\mathbb{F}_a$ ($a=0,1,2,...$)  \cite{oda}. 

\section{Constructing toric $SU(3)$ structures}
\label{sec:toricsu3}
In this section we review and extend the $SU(3)$ structure construction of \cite{Larfors:2010wb}. In addition, we specify the topological restrictions for the existence of toric $SU(3)$ structures.  A recent discussion of some of these topological aspects can be found in \cite{Dabholkar:2013qz}.

\subsection{Topological constraints}\label{sec:su3top}

In the last section we found that all three-dimensional SCTVs admit an $U(3)$ structure, specified by the triplet $(\cM, \widetilde{G}, \widetilde{J}|)$. An $SU(3)$ structure is possible if there exist a pair of nowhere vanishing forms on $\cM$ that satisfy
\be \label{eq:su3cond}
\begin{split}
\Omega\wedge J&=0 \, , \\
\Omega\wedge\overline{\Omega}&=-\frac{4i}{3}J^3\neq 0 \; ,
\end{split}
\ee
where $J$ is a real two-form and $\Omega$ is a complex decomposable three-form. The real two-form $\widetilde{J}|$ must thus be complemented with a nowhere-vanishing three-form if a further reduction of the structure group should take place. This is  a topological restriction on the manifold, which is usually formulated as the requirement that the manifold has vanishing first Chern class $c_1 \equiv c_1(T^{(1,0)} \cM) )$.

However, $c_1$ is not quite a topological quantity, since it depends on the choice of holomorphic tangent bundle and consequently on the choice of almost complex structure. A topological condition that is independent of this choice exists: as long as $c_1$ is even in cohomology, so that the manifold is spin, the SCTV allows an $SU(3)$ structure.\footnote{This existence argument is not restricted to toric varieties (see \eg~\cite{bryant}): any oriented, spin six-manifold allows a reduction of the structure group to $SU(3)$, as can be seen by analysis of the spin bundle. The $SU(3)$ torsion is not specified by this construction. I thank Robert Bryant for explaining this point to me.} Indeed, as we will see below, by changing the almost complex structure, we can set $c_1=0$ as is necessary to allow a nowhere-vanishing three-form \cite{Tomasiello:2007eq,Larfors:2010wb}. This is only possible if $c_1$ is even to start with, a condition that is independent of the almost complex structure. For a toric variety, the total Chern class $c=1+c_1 + c_2 + ...+ c_d$, where $c_i \in \Omega^{2i}(\cM)$, is determined by 
\be
c = \prod_{i=1}^n (1 + D_i) \; ,
\ee
where $D_i$ are the Poincar\'e duals of the divisors $D_i: z^i = 0$. The first Chern class is thus given by the sum
\be
c_1 = \sum_{i=1}^n  D_i \; .
\ee 
A $d$-dimensional toric variety has $s$ linearly independent divisors, corresponding to the linearly independent columns in the $U(1)$ charge matrix $Q$. Consequently, $c_1$ can be expressed as a sum of the linearly independent divisors, and will be even if the coefficients of this sum are all even. Changing the almost complex structure can change the signs of these coefficients, so that they cancel rather than add up, but it cannot change whether they are even or odd.

For the SCTVs classified by Oda, we have, with $a,b,n^a$ as in \eqref{eq:qab}-\eqref{eq:qci}
\begin{itemize}
\item $\mathbb{CP}^3$: 
\be 
c_1 = 4 D_1 \; .
\ee
\item $\mathbb{CP}^2$ bundles over $\mathbb{CP}^1$\\
\be
c_1 = (2+a+b) D_1 + 3 D_5 \; .
\ee
\item $\mathbb{CP}^1$ bundles over two-dimensional SCTVs\\
\be
c_1 = \sum_{a=1}^{n-2} (1+n^a) D_a + 2 D_n \; ,
\ee
where the first sum will be simplified further once the charge components $q^a_i$ for the two-dimensional SCTV are given, since these give the linear relations between the first $n-2$ divisors.
\end{itemize}
From these values, it is immediately clear that $\mathbb{CP}^3$ has an even first Chern class, and that $\mathbb{CP}^2$ bundles over $\mathbb{CP}^1$ always have odd first Chern class.  For $\mathbb{CP}^1$ bundles over two-dimensional SCTVs $c_1$ depends on the twisting parameters $n^a$. It is not difficult to see that one can always choose $n^a$ such that all coefficients are even: let $D_n, D_1, \ldots, D_{s-1}$ be a linearly independent basis of divisors. Then
\be
c_1 = \sum_{a=1}^{n-2} (1+n^a) D_a + 2 D_n = \sum_{a=1}^{s-1}  \left(1+\sum_{i=s}^{n-2} q_i^a + n^a \right) D_a + 2 D_n
\; ,
\ee
and by choosing, say, $n^a = 1+\sum_{i=s}^{n-2} q_i^a$ all coefficients in the sum are even.

In conclusion, of these three types of SCTVs, only two allow $SU(3)$ structures: $\mathbb{CP}^3$ and $\mathbb{CP}^1$ bundles over two-dimensional SCTVs. In the latter case an $SU(3)$ structure is allowed when the twist parameters of the $\mathbb{CP}^1$ bundles are chosen to appropriate values.  Such a choice is always possible.

\subsection{Construction of $J$ and $\Omega$}
\label{sec:su3constr}

If we want to use an SCTV for the purpose of string theory compactifications, it is not enough to know that it permits an $SU(3)$ structure. We also need information about its torsion classes, which are given by the exterior derivatives of $J$ and $\Omega$ \cite{chiossi,Cardoso:2002hd}
\bea 
\label{eq:torsionclass}
d J&=-\frac{3}{2}\mbox{Im}(\mathcal{W}_1\overline{\Omega})+\mathcal{W}_4\wedge J+\mathcal{W}_3 \, , \\ \nn
d \Omega&= \mathcal{W}_1 J\wedge J+\mathcal{W}_2 \wedge J+\mathcal{W}_5 \wedge \Omega ~,
\eea
where $\mathcal{W}_1$ is a function, $\mathcal{W}_2$ is a primitive (1,1)-form and $\mathcal{W}_3$ is a real
primitive $(1,2)+(2,1)$-form. Here, primitivity means that the form contracts to zero with $J$. The Lie forms $\mathcal{W}_4$, $\mathcal{W}_5$ are both real one-forms.\footnote{It is only the (0,1) piece of $\cW_5$ that contributes to \eqref{eq:torsionclass}, so an alternative definition as a complex (1,0)-form is common. Since a real one-form and a complex (1,0)-form carry the same number of degrees of freedom, the two definitions are exchangeable.} For a Calabi--Yau manifold, all torsion classes are zero.

To determine the torsion classes we thus need explicit expressions for $J$ and $\Omega$, which we construct following \cite{Tomasiello:2007eq,Gaiotto:2009yz,Larfors:2010wb}.  As discussed in section \ref{sec:sctv}, we already have a candidate two-form: the inherited K\"ahler form $\widetilde{J}$. In addition, a (3,0)-form (with respect to the inherited complex structure) $\widetilde{\Omega}$ can be constructed on the toric variety by contraction of the holomorphic top form $\Omega_{\mathbb{C}}$ of the ambient space $\mathbb{C}^n$:
\be \label{eq:omtilde}
\widetilde{\Omega}:= \left(\mathrm{det}g_{ab}\right)^{-1/2}\prod_{a=1}^{s}\iota_{V^{a}}
\Omega_{\mathbb{C}} \; .
\ee
Here $V^a$ are the generators of the $U(1)$ action \eqref{holvec} and the factor containing the determinant of \eqref{gab} is needed for normalisation. $\widetilde{\Omega}$ is a vertical form, and its restriction $\widetilde{\Omega}|$ is a regular (without poles) form on the SCTV, with exterior derivative
\be
\d \widetilde{\Omega}| = -\frac{1}{2} \d \ln \left(\mathrm{det}g_{ab}\right) \w \widetilde{\Omega}| \; .
\ee
It is straightforward to show that the pair $(\widetilde{J}|, \widetilde{\Omega}|)$ satisfies the orthogonality and normalisation conditions \eqref{eq:su3cond} (see \cite{Larfors:2010wb} for details). Consequently, the two forms define a local $SU(3)$ structure. 

However, $\widetilde{\Omega}$ is not invariant (it does not have zero $U(1)$ charge)
\be
\mathcal{L}_{\mathrm{Im}V^a} \widetilde{\Omega} = \sum_{i=1}^n Q_i^a \; ,
\ee
and is thus only locally defined. The non-zero charge of $\widetilde{\Omega}$ is linked to the non-vanishing first Chern class, as a globally defined three-form is only allowed when $c_1 = 0$.  As a consequence, the $SU(3)$ structure defined by $(\widetilde{J}|, \widetilde{\Omega}|)$ is only locally defined. To obtain a globally defined three-form we must ``twist'' the SCTV along some divisor so that $c_1$ vanishes. Clearly, this can be accomplished by constructing a twisted three-form with zero $U(1)$ charge, which is possible if there exist a one-form $K$ on $\mathbb{C}^n$ with the following properties:
\begin{enumerate}
\item It is (1,0) (with respect to the inherited complex structure) and vertical: $P(K)=K$.
\item It is an eigenform of $\mathcal{L}_{\mathrm{Im}V^a}$ (\ie{} 
it has definite  $Q^a$-charge):
\be
\mathcal{L}_{\mathrm{Im}V^a}K=q^a K
~,
\ee 
where $q^a$  is half the $Q^a$-charge of ${\Omega}_{\mathbb{C}}$:
\be \label{eq:chargecond}
q^a = \frac{1}{2} \sum_{i=1}^n Q_i^a
~.
\ee 
\item It is nowhere-vanishing, and hence can be normalised to:
\be
K\cdot \bar{K}=2
~,
\ee
where the dot on the left-hand side denotes contraction 
of indices  with respect to the inherited metric $\tilde{G}$. 
\end{enumerate}
Just as $\widetilde{\Omega}$, $K$ is not invariant, and hence only locally defined; consequently it does not restrict the structure group or the topology of the three-fold. With its help we can construct a local $SU(2)$ structure. After normalising $K \cdot \bar{K}=2$,  we define
\be \label{eq:su2def}
\begin{split}
\omega&:= -\frac{i}{2}~\! \bar{K} \cdot \widetilde{\Omega}\vert\\
j&:=\widetilde{J}\vert-\frac{i}{2}K\wedge \bar{K}
\end{split}
\ee
which form a local $SU(2)$ structure. In particular, the $SU(2)$ conditions
\be \label{eq:su2cond}
\begin{split}
\omega\wedge\bar{\omega}&= 2j\wedge j\\
\omega\wedge j&=0
\end{split}
\ee
can be shown to hold from \eqref{eq:su3cond} and \eqref{eq:su2def}. A property that follows from this construction is that  $K$ and $\bar{K}$ contracts to zero with $j$ and $\omega$. The local $SU(3)$ structure is then given by
\be \label{eq:su3local}
\begin{split}
\widetilde{J}\vert&= j + \frac{i}{2}K\wedge \bar{K}\\
\widetilde{\Omega}\vert&=i  K\wedge\omega \; .
\end{split}
\ee

We now perform the ``twist'': a new $SU(3)$ structure is constructed by switching $K \leftrightarrow \bar{K}$ in \eqref{eq:su3local}: 
\be  \label{eq:su3global}
\begin{split}
J&:=\alpha j- \frac{i\beta^2}{2}K\wedge \bar{K}\\
\Omega&:=e^{i\gamma}\alpha\beta  \bar{K} \wedge\omega  \; ,
\end{split}
\ee
where the parameters $\alpha, \beta, \gamma$ are non-zero real functions. Using  \eqref{eq:su2cond} it is straightforward to show that $(J,\Omega)$ satisfy the $SU(3)$ conditions \eqref{eq:su3cond}, and $\Omega$ can also be shown to be complex decomposable for all $\alpha, \beta, \gamma$ \cite{Larfors:2010wb,Larfors:2011zz}. Furthermore, the charge of $J$ and $\Omega$ are both zero by construction, since $Q(\bar{K})= -Q(K) = -Q(\omega)$. Thus, a global $SU(3)$ structure is constructed.

The real functions $\alpha, \beta, \gamma$  in the global $SU(3)$ structure \eqref{eq:su3global} are not constrained by \eqref{eq:su3cond} (but, in a string vacuum, they will be restricted by supersymmetry constraints and the equations of motion).  Two limits in the parameter space are of particular interest, namely
\be
\alpha = -\beta^2 \; \; , \; \; \beta=1\; \mbox{ , and }
\alpha = +\beta^2 \; \; , \; \; \beta=1 \; .
\ee
In the first limit the real two-form $J =- \widetilde{J}|$ is closed, and the $SU(3)$ structure is symplectic. In the second limit, the metric defined by $(J,\Omega)$ equals the metric $\tilde{G}$ induced from the canonical metric on $\mathbb{C}^n$ \cite{Larfors:2010wb}.

\subsection{Existence of $K$}
 \label{sec:Kreq}

At this point it should be clear that given a one-form $K$ we can explicitly construct an $SU(3)$ structure. What remains to show is when such a one-form exists. We have seen above that an $SU(3)$ structure is always allowed once the first Chern class, $c_1$ is even in cohomology. We will now show that if this constraint is satisfied, the one-form $K$ can always be constructed.

As discussed in section \ref{sec:su3constr}, there are three conditions on $K$. First, it should be $(1,0)$ with respect to $\widetilde{\mathcal{I}}$ and vertical. These conditions are met by linear combinations of $\cD z^i$ that are also eigenvectors with eigenvalue 1 of the projection matrix $P_{ij}$:  
\be
K_i P_{ij}= K_j~.
\ee
$P_{ij}$ has rank three, so there are three such linearly independent eigenvectors. For $\mathbb{CP}^3$ they can be taken as 
\be \label{eq:evcp3}
K_1 = (-z^2,z^1,0,0); \quad
K_2 = (0,0,-z^4,z^3); \quad
K_3 = (-z^4,0,0,z^1) ~,
\ee
where the first vector corresponds to the form $K_1 = -z^2 \cD z^1 + z^1 \cD z^2$ etc. For $\mathbb{CP}^1$ bundles, with coordinates $z^{n-1}, z^n$ along the $\mathbb{CP}^1$ fibre, the eigenvectors have a similar form, and we list them for bundles with up to six generators in table \ref{tab:1} in section \ref{sec:Kuniq}. Schematically, there are two eigenvectors with zero components along the $\mathbb{CP}^1$ fibre, and one with non-zero components:
\be \label{eq:evcp1bdl}
K_1 = (*,\ldots,*,0,0); \quad
K_2 = (*,\ldots,*,0,0); \quad
K_3 = (*,\ldots,*,*) ~,
\ee
where $*$ means that the entry is not necessarily zero. As long as the $\mathbb{CP}^1$ fibration is non-trivial (\ie~non-zero twist parameters $n^a$), there is no eigenvector of the form $(0,\ldots,0,*,*)$.

Thus, the first condition on $K$ can always be fulfilled. The second condition is that $K$ should have half the $U(1)$ charge of $\widetilde{\Omega}$. It is easy to see that this can only be satisfied when the charge of $\widetilde{\Omega}$ is even (since no function of the $z^i$ has fractional charge). This will restrict the twist parameters $n^a$ in $Q$, just as the condition on the first Chern class did. In fact, the even charge condition on $\widetilde{\Omega}$ exactly corresponds to requiring that $c_1$ is even in cohomology, and so can be solved for $\mathbb{CP}^3$ and all $\mathbb{CP}^1$ fibrations. Concretely, once the $n^a$ are fixed, we read off the charge of $K_{i}$, and look for functions $\alpha_{i}$ so that 
\be
\hat{K} =\sum_{i=1}^3 \alpha_i K_i
\ee
has the required charge. Such functions $\alpha_{i}$ can always be found, and several consistent choices may exist. 

Thirdly, we must check that the norm of $\hat{K}$ is nowhere-vanishing, so that the twisted $SU(3)$ structure is well-defined. This is possible by choosing $\alpha_i$ so that $|\hat{K}|^2$ is bounded from below by a positive combination of the K\"ahler moduli (recall that these are positive for non-singular varieties). This step requires a bit more work than the charge condition; in particular the K\"ahler and Mori cones of the variety needs to be identified as described in appendix \ref{sec:Mori}.  Again, several consistent choices may exist and we will come back to the question of uniqueness in the examples. 

To conclude, a one-form $K$ fulfilling the three conditions can always be found if $c_1$ is even in cohomology. Consequently, an $SU(3)$ structure can be constructed and its torsion classes can be computed.

\section{Properties of toric $SU(3)$ structures}

In the last section, we showed that SCTVs with even first Chern class allow $SU(3)$ structures, and we also constructed the defining forms $J$ and $\Omega$. To investigate if these structures are relevant for string vacua, we need to further analyse the associated metric and torsion classes. In this section we give explicit forms of the almost complex structure and metric defined by the $SU(3)$ structure, and use metric positivity to derive constraints on the parameters of the construction. We also compute generic properties of the torsion classes of toric varieties, particularly focusing on how they are affected by changes of the parameters.

\subsection{Almost complex structure and metric}
\label{sec:metric}
Any $SU(3)$ structure has a metric, that is determined by $(J,\Omega)$  as follows \cite{Hitchin:2000jd}. First, $\Omega$ specifies an almost complex structure by
\be
\mathcal{I} = \frac{\widehat{\mathcal{I}}}{\sqrt{-\text{tr}\, \frac{1}{6}\,\widehat{\mathcal{I}}^2}}
\mbox{ , where }~
\widehat{\mathcal{I}}_k{}^l = \varepsilon^{lm_1\dots m_5} (\mbox{Re} \Omega)_{km_1m_2} (\mbox{Re} \Omega)_{m_3m_4m_5} 
\ee
and $\varepsilon^{m_1\dots m_6}=\pm1$ is the totally antisymmetric symbol in six dimensions. Complex decomposability of $\Omega$ guarantees that $\mathcal{I}^2=-1$. Using $\mathcal{I}$ and $J$, the metric is then given by 
\be
\label{eq:su3metric}
G_{mn}=-\mathcal{I}_m{}^lJ_{ln}~.
\ee
The construction does not guarantee that the metric is positive definite.

For the local $SU(3)$ structure $(\widetilde{J},\widetilde{\Omega})$ we can thus compute, using \eqref{eq:su3local},
\be \label{eq:ifs}
\widehat{\widetilde{\mathcal{I}}}_k{}^l = \frac{3}{2} \varepsilon^{lm_1\dots m_5} \mbox{Re} \left(
(K_k \omega_{m_1m_2} - 2 K_{m_1} \omega_{k m_2})
\bar{K}_{m_3} \bar{\omega}_{m_4 m_5}
\right)
\ee
where further terms vanish due to index antisymmetrisation. It can be checked that this is just the inherited complex structure from $\mathbb{C}^n$, and the associated metric is the inherited metric $\widetilde{G}$  (this is also known as the Fubini--Study metric on $\mathbb{CP}^3$):
\be
\widetilde{G}_{mn}=-\widetilde{\mathcal{I}}_m{}^l\widetilde{J}|_{ln}~.
\ee

For the global $SU(3)$ a similar computation yields
\bea \label{eq:isu3}
\widehat{\mathcal{I}}_k{}^l &= \frac{3}{2}\alpha^2 \beta^2 \varepsilon^{lm_1\dots m_5} \mbox{Re} \left(
(-K_k \omega_{m_1m_2} - 2 K_{m_1} \omega_{k m_2})
\bar{K}_{m_3} \bar{\omega}_{m_4 m_5}
\right) \; .
\eea
Note that the phase of $\Omega$ (i.e. $\gamma$) does not affect $\mathcal{I}$, and the factors of $\alpha$ and $\beta$ will cancel in the normalised almost complex structure $\mathcal{I}$. Thus, the only difference between this almost complex structure and the inherited complex structure \eqref{eq:ifs} is a relative sign. This sign reflects the twisting of the toric variety, and matches the relative sign found in the almost complex structures of twistor spaces, see equations (3.5) and (3.6) in \cite{Tomasiello:2007eq}. Using \eqref{eq:ifs} and \eqref{eq:isu3} it is straightforward to show 
\bea
\widetilde{\mathcal{I}}_k{}^l K_l &= i  K_k = -\mathcal{I}_{ k}{}^l K_l \\ \nn
\widetilde{\mathcal{I}}_k{}^l \omega_{lm} &= i  \omega_{km} =  \mathcal{I}_{ k}{}^l \omega_{lm}
 \; .
\eea
Consequently, $\widetilde{\Omega}$ and $\Omega$ are (3,0) with respect to their associated almost complex structures, as is required for the consistency of the construction.

The metric associated to the $SU(3)$ structure is given by inserting $J$ and \eqref{eq:isu3} in \eqref{eq:su3metric}. A short computation gives 
\be \label{eq:su3metric2}
G_{mn} =\alpha \left[ \widetilde{G}_{mn} + \left(\frac{\beta^2}{\alpha}-1\right) \mbox{Re} \left(K_m \bar{K}_{n}
\right) \right]
\; .
\ee
This expression for the metric is another manifestation of the twisting of the toric variety by $K$.  In the parameter limit $\alpha = \beta^2 = 1$ it simplifies to the induced metric. Thus, contractions are greatly simplified in this limit, which is helpful when computing the torsion classes.

Expressing the $SU(3)$ structure metric as in \eqref{eq:su3metric2} facilitates the check of metric positivity: $G$ is positive definite if for any non-zero vector $\mathbf{v}$
\be
0 < \mathbf{v}^{T} G \mathbf{v} = \alpha \left[ \mathbf{v}^{T}\widetilde{G}\mathbf{v} + \left(\frac{\beta^2}{\alpha}-1\right) \mathbf{v}^{T} \mbox{Re} \left(K \bar{K} \right) \mathbf{v} \right] \; .
\ee
$K$ is directed along a certain direction in the space of one-forms, and so $\mbox{Re} \left(K \bar{K}\right)$ will contribute to a block of $G$. As a consequence,  some of the eigenvalues $G$ will be proportional to those of $\widetilde{G}$, with proportionality coefficient $\alpha$. Since $\widetilde{G}$ is positive definite, the condition
\be \label{eq:parpos1}
\alpha > 0 \; , 
\ee
is thus necessary for metric positivity.\footnote{This argument can be rephrased using Sylvester's criterion \cite{gilbert}.} This is a severe restriction on the parameters, and it shows that the $SU(3)$ structure does not have a positive definite metric in the symplectic limit $\alpha = -\beta^2 =-1$. Further parametric constraints can be derived once the properties of $\mbox{Re} \left(K \bar{K}\right)$ are known. For example, if $\mbox{Re} \left(K \bar{K}\right)$ is positive semidefinite, $\beta^2 \ge \alpha$ is a sufficient (but not necessary) condition for  metric positivity.

\subsection{Torsion classes, choices of $K$ and parameters}
\label{sec:torsion}
The torsion classes of a toric $SU(3)$ structure are determined by the exterior derivatives of $K$, $\omega$ and the parameters $\a, \b, \g$. In this section we discuss their general properties. Let us first note that the parametric freedom given by $\a, \b, \g$ is a great help when computing the torsion classes. Contractions are needed in order to decompose $\d J$ and $\d \Omega$ in $SU(3)$ representations, and since the metric \eqref{eq:su3metric2} tends to be complicated for generic choices of $K$, these are computationally expensive. It is therefore very useful that parameter choices exist where either the metric simplifies, or some torsion classes are set to zero. 

For constant parameters $\a, \b, \g$, the torsion classes are uniquely determined by the exterior derivatives of $K$ and $\omega$. Decomposing $J$ as
\be \label{eq:Jdec}
J = \alpha \widetilde{J}| + \frac{i}{2} (\a + \b^2) K \w \bar{K}
\ee
shows that 
\be
\d J |_{\d\a=\d\b=0} = -(\a + \b^2) \mbox{Im} (\d K \w \bar{K}) \; .
\ee
Consequently, up to contributions from $\d \a$ and $\d\b$, the torsion classes $\cW_1$, $\cW_3$ and $\cW_4$ are completely determined by $\d K$. If, as we will see in some examples,
\be \label{eq:dKgen}
\d K = \delta \omega + \Psi \w K
\ee
where $\delta$ is a function and $\Psi$ a one-form, we find that $\cW_1 \propto \delta$ and
\be
\begin{split}
\cW_4 \w J + \cW_3 = i (\a + \b^2) \mbox{Re} (\Psi) \w K \w \bar{K} 
\; ,
\end{split}
\ee
since $K \w \bar{K}$ is imaginary. Evidently, if $\mbox{Re} (\Psi)$ is zero, so are $\cW_3$ and $\cW_4$. 

As is clear from \eqref{eq:Jdec}, $J$ is a closed form in the limit $\alpha = -\beta^2, \beta=1$. Thus the only non-zero torsion classes in this limit are $\cW_2$ and $\cW_5$, and the $SU(3)$ structure is symplectic. Moving away from this limit generically switches on all torsion classes. As an example, the primitivity condition on $\cW_2$ 
\be
\cW_2 \w J \w J = 0
\ee
depends on $\alpha$ and $\beta$. Thus, $\cW_2$ computed in one parameter limit will give contributions to both $\cW_1$ and $\cW_2$ in a different region of parameter space. Another relevant observation is that the phase of $\cW_1$ and $\cW_2$ is completely determined by $\gamma$.

Finally, it was shown in \cite{Gray:2012md} that non-constant $\alpha, \beta, \gamma$ contribute additional terms to $\cW_3, \cW_4$ and $\cW_5$. In summary, we have
\be
\begin{split}
\label{sctvtorsions}
&\cW_1= (\alpha+\beta^2) e^{i \g} \cW_1^0 \\
&\cW_2 =e^{i\g} \cW_2^0 \\
&\cW_3 = (\alpha+\beta^2) \cW_3^0   
 + \left(\chi - \frac{1}{4} (J \lrcorner \chi )\wedge J \right)
\\
&\cW_4 = (\alpha+\beta^2) \cW_4^0 + \frac{1}{4} J \lrcorner \chi
\\
&\cW_{5}=\cW_5^0 + \d  \ln (\alpha \beta) + \mathcal{I} \d  \gamma  ~,
\end{split}
\ee
where $\lrcorner$ denotes contraction, $\cW_i^0$ a reference value for the torsion class (computed with constant $\a, \b, \g$), $\mathcal{I} \d = i (\partial - \bar{\partial})$ and we recall that $\cW_5$ is real in our conventions.  The three-form $\chi$ that contributes to $\cW_3$ and $\cW_4$ is given by
\be
\chi =\d \ln \alpha \wedge J  + i \frac{\beta^2}{2} \d (\ln \alpha -2 \ln \beta)  \wedge K \wedge \bar{K} \ .
\ee
When $\alpha \propto \beta^2$, with constant proportionality coefficient, we have $\chi = \d \ln \alpha \wedge J$. This lacks a primitive piece, and so does not contribute to $\cW_3$, but adds an exact term to $\cW_4$.

From the above reasoning, it is clear that exact contributions to $\cW_4$ and $\cW_5$ can be compensated by parameter choices. This phenomenon is related to an observation by Chiossi and Salamon \cite{chiossi}: it follows from the second $SU(3)$ condition in \eqref{eq:su3cond} that under conformal transformations $g \to e^{f} g$, where $f$ is any real function, the torsion classes $\cW_4$ and $\cW_5$ both transform by the addition of exact pieces. Thus, if $\cW_4$ and $\cW_5$ are exact and $3\cW_5-2\cW_4= 0$ we can make a conformal transformation to an $SU(3)$ structure with vanishing Lie forms.

Since $\alpha$ and $\beta$ are two real functions, the parametric freedom  is a bit larger than conformal transformations: as long as $\cW_4^0$ and $\cW_5^0$ are exact and proportional, with constant coefficient of proportionality, they can be set to zero. Clearly, if for some function $p$ and constant $A$
\be \label{eq:dpcontr}
\cW_4^0 =\frac{1}{\b^2} \d \ln p \; , \; 
\cW_5^0 = A \d \ln p \; ,
\ee
we can choose $\alpha = \frac{2A-3}{3}\beta^2$ and $\beta=p^{-A/3}$ to set $\cW_4 = 0 = \cW_5$. If $\cW_{4,5}^0$ are exact but have a non-constant quotient, only one linear combination of them can be set to zero.

In a given example, there may be additional parameter limits that set other sets of torsion classes to zero. In general, care is needed when distinguishing the $SU(3)$ structures that are obtained through the construction in section \ref{sec:su3constr}, as it is possible that different choices of $K$ lead to $SU(3)$ structures that are equivalent up to changes in the parameters $\alpha, \beta$ and $\gamma$. We will discuss this phenomenon in explicit examples in the following section, and it would certainly be interesting to study this question in more depth in the future.

\section{Examples} \label{sec:Kuniq}

In section \ref{sec:toricsu3}, we argued that whenever $c_1$ is even in cohomology, there exist a one-form $K$ fulfilling the three requirements discussed in section \ref{sec:su3constr}. The choice of $K$ is not unique, as was first pointed out in \cite{Larfors:2010wb}, which leads to the possibility of having multiple $SU(3)$ structures on a single toric variety. In this section, we construct $K$ for $\mathbb{CP}^3$ and toric $\mathbb{CP}^1$ bundles over two-dimensional SVTVs with up to six generators. For  $\mathbb{CP}^3$ and $\mathbb{CP}^1 \hookrightarrow \mathbb{F}_0$, we show that simple changes to $K$ lead to parametrically inequivalent $SU(3)$ structures on example manifolds.\footnote{The symbolic computer program \cite{bonanos} has been used for all explicit computations of torsion classes.} In addition, we check if the $SU(3)$ structures thus obtained lead to vacua of the type discussed in section \ref{sec:survey}.  

\subsection{$\mathbb{CP}^3$}

Our first example, $\mathbb{CP}^3$, has been studied at length in the literature, starting with the classical papers \cite{Nilsson:1984bj,Sorokin:1985ap} to more modern studies \cite{Behrndt:2004km,Behrndt:2004mj,Lust:2004ig,Tomasiello:2007eq,Koerber:2008rx}. In the symplectic quotient description, this manifold can be constructed as a subspace of $\mathbb{C}^4$ using \eqref{kmoddef} and the single charge
\be
\label{eq:cha}
Q^1=\begin{pmatrix} 1&1&1&1 \end{pmatrix} ~.
\ee
The local $SU(3)$ structure is given by
\be
\widetilde{J} = \sum_{i=1}^4 \cD z^i \wedge \cD \bar{z}^i \; , \; 
\widetilde{\Omega} =
\frac{1}{\sqrt{\mbox{det}g_{ab}}}
\left( z^1 \cD z^{234} - z^2 \cD z^{134} + z^3 \cD z^{124} -z^4 \cD z^{123} \right) \; ,
\ee
where the prefactor is a positive constant (see \eqref{gab})
\be \label{eq:p3xi}
\mbox{det}g_{ab} = \sum_{i=1}^4 |z_i|^2 = \xi > 0 \; .
\ee
Here $\xi$ is the (coordinate independent) K\"ahler modulus of $\mathbb{CP}^3$, which is strictly positive in the K\"ahler cone. Consequently, $\widetilde{\Omega}$ is a closed form.

The vertical one-form $K = \sum_{i=1}^3 \alpha_i K_i$ is a linear combination of the $P_{ij}$ eigenvectors \eqref{eq:evcp3}. 
Since $Q(K_i)=2$ already is half of that of $\widetilde{\Omega}$, we must choose $\alpha_i$ with charge 0. The choice of $\alpha_i$ completely determines the $SU(3)$ structure, and different choices will lead to different torsion classes.  

\subsubsection*{Half-flat $SU(3)$ structure}
An interesting choice for $\hat{K}$ is 
\be \label{eq:therightK}
\hat{K}=(-z^2,z^1,-z^4,z^3) \; ,
\ee
to be read as a vector in the $\cD z^i$ basis. With respect to the Fubini--Study metric, this has constant norm  $|\hat{K}|^2 =\mbox{det}g_{ab}=\xi \neq 0$, and so we can define
\be
K = \frac{1}{\sqrt{\mbox{det}g_{ab}}} (-z^2,z^1,-z^4,z^3) \; ,
\ee
which has norm 2 and can be used to construct the global $SU(3)$ structure \eqref{eq:su3global}.

It can be checked that this $K$ gives a positive semidefinite contribution to the $SU(3)$ metric \eqref{eq:su3metric2}, and so a sufficient condition for  positive definiteness of the latter is $\alpha>0$ and $\beta^2>\alpha$. On closer inspection, it can be shown that the last of these conditions is superfluous, and that metric positivity is guaranteed by only imposing the first constraint.

The torsion classes are straightforward to compute. First, we note that $\d K$ is proportional to $\omega$:
\be
\d K = \frac{2}{\sqrt{\mbox{det}g_{ab}}} \omega \; ,
\ee
and that we can fix $\gamma$ so that $\d \Omega$ is real (or imaginary). The first assertion sets $\cW_3 = 0 = \cW_4$, while the second implies that $\cW_5=0$. $\cW_2$ is non-zero and can be computed by contracting $J$ with $\d \Omega$ in the limit $\alpha=\beta^2$, and then using the result as an ansatz for general parameters. The result, for constant $\a, \b, \g$, is
\be
\begin{split}
\cW_1 &= \frac{ 4 e^{i \gamma} (\alpha+\beta^2)}{3 \alpha \beta \sqrt{\mbox{det}g_{ab}}}  \\
\cW_2 &=  \cW_1 \frac{2\beta^2-\alpha}{\alpha+\beta^2} \left(
J +\frac{3i \beta^2}{2} K \w \bar{K}
 \right) \\
\cW_3 &= 0 \; ,
\cW_4 = 0 \; ,
\cW_5 = 0\; .
\end{split}
\ee
Since only $\cW_1$ and $\cW_2$ are non-zero, this is an example of a restricted half-flat $SU(3)$ structure. In fact, we have reproduced the $SU(3)$ structure found from a twistor analysis in \cite{Tomasiello:2007eq} and a coset perspective in \cite{Koerber:2008rx}. Comparing with section \ref{sec:survey}, it is straightforward to check that this structure satisfies necessary requirements for several string vacua, such as the type IIA $\cN=0,1$ AdS vacua and the calibrated $\cN=0$ vacua of either type IIB (with O5 planes) or heterotic string theory. To fully investigate that all constraints are satisfied for a particular vacuum goes beyond the scope of this paper, and we refer the reader to \eg~\cite{Tomasiello:2007eq,Koerber:2008rx,Caviezel:2008tf} for a more detailed discussion.  

\subsubsection*{Modified $SU(3)$ structure}
Let us now investigate if there are different choices of $\alpha_{1,2}$, such that the new $K$ still fulfils the verticality, charge and norm conditions. We focus on the last condition, which is most constraining. On $\mathbb{CP}^3$, there is only one $U(1)$ charge, which says that \eqref{eq:p3xi} is non-zero. However, adding any non-negative combination of $|z^i|^2$ to $\xi$ also gives a nowhere-vanishing expression. Since $K_1$ and $K_2$ are orthogonal vectors with non-negative norm, we can thus change $\alpha_{1,2}$ and get a new $K$ with nowhere vanishing norm.  For simplicity, we take $\alpha_{1,2}$ to be real constants; we will comment on non-constant functions below. 

The norm of the new form $\hat{K}_{new}$ is non-constant  for $\alpha_{1} \neq \alpha_2$
\be 
p = |\hat{K}_{new}|^2=\a_1 \mbox{det}g_{ab} + (\a_2-\a_1) |K_2|^2 \; .
\ee
The one-form $K_{new} = 1/\sqrt{p} \hat{K}_{new}$ gives a positive semidefinite contribution to the $SU(3)$ metric \eqref{eq:su3metric2}. Again, it can be shown that metric positivity only requires $\alpha>0$. 
 
The exterior derivative of $K$ is no longer proportional to $\omega$, and in particular gives a term $-\frac{1}{2} \d \ln p \w K$ which will contribute to the Lie forms.  Thus, after a straightforward computation, we find 
\be
\begin{split}
\cW_1^0 &= \frac{ 4 \a_1 \a_2   \sqrt{\mbox{det}g_{ab}}}{3 \alpha \beta p }  \\
\cW_2^0 &=   (2\beta^2-\alpha)\cW_1^0 \left(
J +\frac{3i \beta^2}{2} K \w \bar{K}
 \right) \\
\cW_3^0 &= - \cW_4^0 \wedge \left( J + i \beta^2 K \w \bar{K} \right) \; , \\
\cW_4^0 &= \frac{1}{2 \beta^2} \d \ln  p \; , \\
\cW_5^0 &= 2 \d \ln  p \; ,
\end{split}
\ee
which should be inserted into \eqref{sctvtorsions} to get the torsion classes for general parameters. As is clear from these equations, the effect of choosing non-trivial $\alpha_{1,2}$ is that $\cW_1$ and $\cW_2$ are rescaled, $\cW_3, \cW_4$ and $\cW_5$ are all non-zero and $\cW_4$ and $\cW_5$ are exact. Even though these changes can largely be compensated by a change in the parameters $\a, \b, \g$, no choice of these parameters take us back to the restricted half-flat $SU(3)$ studied in the previous subsection. There are, however, several parametric limits where some of the torsion classes are zero. In fact, with different parametric choices, we can turn on or off all torsion classes but $\cW_1$ (this is only zero in the symplectic limit $\alpha = -\beta^2$ which is excluded by metric positivity). Comparing with table \ref{tab:su3N1} and \ref{tab:su3N0}, we note that no $\cN=1$ vacuum can be constructed using this $SU(3)$ structure, but that $\cN=0$ vacua of type IIB (with O5 planes) and heterotic string theory may be allowed. Again, we leave a detailed investigation to the future.

We thus conclude that in the toric formulation, $\mathbb{CP}^3$ allows a more general $SU(3)$ structure than has been found through twistor space and coset studies. To further stress this point we can allow the $\a_i$ to be non-constant. This does not change the metric, but will in general change all the torsion classes. Most importantly, $\cW_4$ and $\cW_5$ are no longer exact, and so one of the necessary constraints for string vacua cannot be met. The connection between $\cW_3$ and $\cW_4$ is also lost. All this can be understood at the level of $dK$: for generic non-constant $\a_i$, the relation \eqref{eq:dKgen} fails since $\d \a_1$ and $\d \a_2$ need not be equal.

\begin{table}
\hspace{-0.5cm}
\begin{tabular}[htb]{| l | l | p{11.5cm} |}
\hline
$N$ & $q$ & $K_i$\\
\hline
\hline
3
 & $\begin{pmatrix} 
1&1&1 
\end{pmatrix}$
 & 
  $\begin{matrix}
&K_1=  (&-z^3,& 0,&  z^1,&  0,& 0 ) \\
&K_2= (&0,& -z^3,& z^2,& 0,& 0 ) \\
&K_3=  (&c^1z^{45},& 0,& 0,& -z^{14},& z^{15} )
 \end{matrix}$ \\
\hline
\hline
4 &
 $\begin{pmatrix}
0&1 & 0 &1\\
1&a & 1 &0 
\end{pmatrix}$
& 
 $\begin{matrix}
&K_1=&(-z^3\;, \; 0 \; \;, \; z^1&, &0 \;, &0 \;, &0 \;)& &\\ 
&K_2=&(a z^{24},\; -z^{14} \;, \; 0 \;&,  &z^{12},  &0 \;, &0 \;)& &\\
&K_3=&(\;[c^2-a c^1]z^{256}&,  &c^1 z^{156}, &0, &0, &-z^{126}, &z^{125})
 \end{matrix}$
 \\
\hline
\noalign{\smallskip}
\hline
5 
 & 
 $\begin{pmatrix}
1&a & 1 &0 &0\\ 
0&1 & 0 & 0 &1\\
1&a+1 & 0 &1 &0 
\end{pmatrix}$
& 
$\begin{matrix}
&K_1=& (-z^{34},\;0 \;, \; z^{14} \; ,& z^{13},\; 0,\; 0,\; 0) \;\; & \; & \\
&K_2=& (0,-z^{345}, \; a z^{245},& [1+a]z^{235},\; z^{234}, 0, 0)&  \; &\\
&K_3=& ([c_1+c_3]z^{4567},\; 0,& 0, \; -c_1 z^{1567}, c_2 z^{1467},& \; -z^{1457},& \;z^{1456}) 
\end{matrix}$\\
\hline
\hline
6 
 & $\begin{pmatrix}
-1&1 & -1 &0&0 &0\\ 
1&0&a & 1 & 0 &0\\
0&0&1 & 0 &0 &1\\ 
1&0&a+1 & 0 &1 &0\end{pmatrix}$
& 
$\begin{matrix}
&\,K_1= (z^{245},& z^{145},&  0,&  -z^{125},& -z^{124},& 0,&0,0) \end{matrix}$
 $\begin{matrix}
&K_2= (0,& z^{3456},&   z^{2456},&  -a z^{2356},&  -[a+1]z^{2346},& -z^{2345},&0,0)
\end{matrix}$  
\\
\hline
\hline
6 
 & $\begin{pmatrix}
1&a & 1 &0&0 &0\\ 
2&2a+1&0 & 1 & 0 &0\\
1&a+1&0 & 0 &1 &0\\ 
0&1&0 & 0 &0 &1  
\end{pmatrix}$
&  $\begin{matrix}
&\,K_1= (z^{345},& 0,&  -z^{145},& - 2z^{135},& - z^{134},& 0,&0,0) \end{matrix}$
 $\begin{matrix}
&K_2= (0,& z^{3456},&   -az^{2456},&  -[2a+1] z^{2356},&  -[a+1]z^{2346},& -z^{2345},&0,0)
\end{matrix}$  \\
\hline
\hline
6 
 & $\begin{pmatrix}
1&a & 1 &0&0 &0\\ 
1&a+1&0 & 1 & 0 &0\\
1&a+2&0 & 0 &1 &0\\ 
0&1&0 & 0 &0 &1 
\end{pmatrix}$ 
&  $\begin{matrix}
&\,K_1= (z^{345},& 0,&  -z^{145},& - z^{135},& - z^{134},& 0,&0,0) \end{matrix}$
 $\begin{matrix}
&K_2= (0,& z^{3456},&   -az^{2456},&  -[a+1] z^{2356},&  -[a+2]z^{2346},& -z^{2345},&0,0)
\end{matrix}$   \\
\hline
\hline
\end{tabular}
\caption{\it Charges and vertical one-forms for $\mathbb{CP}^1$ bundles over two-dimensional SCTVs with $N$ generators. $q$ is the charge matrix for the two-dimensional SCTVs, and the charge matrix $Q$ for the $\mathbb{CP}^1$ bundle is given by \eqref{eq:qci}. The one-forms $K_i = K_{i,m} \cD z^m$ are eigenvectors of $P_{ij}$, and the abbreviation $z^{ij..}=z^i z^j..$ is used. For $N=6$, only two of the three linearly independent eigenvectors have been computed. \label{tab:1}}  
\end{table}

\subsection{$\mathbb{CP}^1$ bundles over two-dimensional SCTVs}

Toric $\mathbb{CP}^1$ bundles differ from $\mathbb{CP}^3$ in two important respects. First, the determinant of the symmetric matrix $g_{ab}$ \eqref{gab} is no longer constant. Consequently, the local three-form $\widetilde{\Omega}$ is no longer closed. Second, these varieties are specified by more than one moment map, which can all be used to build up a nowhere-vanishing norm of $K$. This leads to more freedom in the construction, and it is not expected that $K$ should be unique. 

In this section, we first discuss $\mathbb{CP}^1$ bundles over the Hirzebruch surface $\mathbb{F}_0 = \mathbb{CP}^1 \times \mathbb{CP}^1$. This example was first studied in \cite{Larfors:2010wb} where an $SU(3)$ structure was constructed and some of the torsion classes were computed (see also \cite{Gray:2012md}). Here we compute all torsion classes and also discuss how they are affected by changes in the choice of $K$. Secondly, we present valid choices for $K$ on  $\mathbb{CP}^1$ bundles over two-dimensional SCTVs with 3, 5, and 6 generators. This includes the flag manifold $\mathbb{CP}^1$ over $\mathbb{CP}^2$, which is known to allow the same type of half-flat $SU(3)$ structure that $\mathbb{CP}^3$ does \cite{Tomasiello:2007eq,Koerber:2008rx}. 

\subsubsection{$\mathbb{CP}^1$ bundles over $\mathbb{F}_0$}

The charges for a $\mathbb{CP}^1$ fibration over $\mathbb{F}_a$ are
\be
\begin{split}
\label{eq:Qn4}
Q^1&=(0,1,0,1,n^1,0)  \\ 
Q^2&=(1,a,1,0,0,-n^2)  \\ 
Q^3&=(0,0,0,0,1,1) ~.
\end{split}
\ee
As discussed in section \ref{sec:su3top}, $\mathbb{CP}^1$ fibrations only allow $SU(3)$ structures for certain values of the parameters $n^a$. In this example, $n^1$ and $a-n^2$ must be even to obtain an even first Chern class, or equivalently an even $U(1)$ charge of $\widetilde{\Omega}$
\be
Q (\widetilde{\Omega}) = \left(2 + n^1 , 2+a-n^2, 2 \right) \; . \;
\ee
For concreteness, we set $a=0$ from now on, referring to \cite{Larfors:2010wb} for a discussion of non-zero $a$.

The choice of basis for the generators of the $U(1)^3$ group is connected to the value of the parameters, and for the choice \eqref{eq:Qn4} the K\"ahler cone is given by $\widetilde{\xi}^a >0$ only for negative $n^1$ and positive $n^2$. Here $\widetilde{\xi}^a$ are the K\"ahler moduli  that enter the moment maps 
\be \label{eq:mmaps}
\begin{split}
|z^2|^2+|z^4|^2+n^1|z^5|^2&=\widetilde{\xi}^1\\ 
|z^1|^2+|z^3|^2-n^2|z^6|^2&=\widetilde{\xi}^2\\ 
|z^5|^2+|z^6|^2&=\widetilde{\xi}^3 \ .
\end{split}
\ee
We expand upon this issue in appendix \ref{sec:Mori} (see also \cite{Larfors:2010wb} and \cite{Denef:2008wq}). 

We choose $\hat{K}$ as a linear combination of the $P_{ij}$ eigenvectors $K_i$ listed in the second row of table \ref{tab:1}. These three-forms have different $U(1)$ charges, none of which is half of that of $\widetilde{\Omega}$: 
\be
Q (K_1) = \left(0, 2,  0 \right) \; ,\;
Q (K_2) = \left(2, 0, 0 \right)\; ,\;
Q (K_3) = \left(1, 1+a, 2 \right) \; .
\ee
Noting that the parameters $n^a$ can be used to tune the first two components of  $Q (\widetilde{\Omega})$, but not the last, we restrict our ansatz to
\be \label{eq:hatK}
\hat{K} = \alpha_1 K_1 + \alpha_2 K_2 \; .
\ee
For $a=0$, $K_1$ and $K_2$ are orthogonal and (since $n^1 \le 0$ and $n^2 \ge 0$)
\be
|K_1|^2 = |z^1|^2+|z^3|^2 \ge \widetilde{\xi}^2 > 0 \; , \; 
|K_2|^2 = |z^2|^2+|z^4|^2\ge \widetilde{\xi}^1 > 0\; .
\ee
Hence $\hat{K}$ has nowhere vanishing norm if we pick $\alpha_{1,2}$ that cannot be simultaneously zero. 

Turning to the charge condition, we find that $K$ has half the charge of $\widetilde{\Omega}$ if
\be
Q(\alpha_1) = \frac{1}{2}(2+n^1, -2-n^2, 2) \; , \; 
Q(\alpha_2) = \frac{1}{2}(-2+n^1, 2-n^2, 2) \; .
\ee
The simplest solution to these constraints is $\alpha_1 = z^6 \; , \;  \alpha_2=z^5$,
which satisfies the charge condition if we impose $n^1 = -2 \; , \; n^2 = 2$. This choice of $\alpha_i$ was studied in \cite{Gray:2012md,Larfors:2010wb}, where all torsion classes but $\cW_2$ were computed. Using our improved understanding of the $SU(3)$ metric we can now compute this torsion class. In addition, we generalise the choice of $K$ to 
\be \label{eq:choice1}
\alpha_1 = B_1 z^6 \; , \;  \alpha_2=B_2 z^5
\ee
where $B_i$ are real and constant. Non-constant $B_i$ lead to the same changes of the torsion classes as for $\mathbb{CP}^3$; they are all non-zero and the Lie forms are not exact.

We now insert $K = 1/\sqrt{p} \hat{K}$, where $p=|\hat{K}|^2$, in \eqref{eq:su3global} to get the $SU(3)$ structure. The contribution of this $K$ to the $SU(3)$ metric is positive semidefinite; in addition to four zero eigenvalues, the matrix $\mbox{Re}(K \bar{K})$ has two equal positive eigenvalues $E_{K\bar{K}}$, that in the patch $z_1, z_4, z_6 \neq 0$ are given by
\be
\begin{split}
E_{K\bar{K}} = \frac{1}{|z^{14}|^2 p} \Big[
&B_2^2 |z^{15}|^2 |K_2|^2 +
B_1^2 |z^{46}|^2 |K_1|^2 +\\
&4 |z^5|^2 (B_2^2 |z^{125}|^2 + B_1^2 |z^{346}|^2 + 2 B_1 B_2 \mbox{Re} (\bar{z}^{136} z^{245})
\Big] \\
\ge \frac{1}{|z^{14}|^2 p} \Big[
&B_2^2 |z^{15}|^2 |K_2|^2 +
B_1^2 |z^{46}|^2 |K_1|^2 +
4 |z^5|^2 ( B_2 |z^{125}| - B_1 |z^{346}|)^2 \Big]
\; ,
\end{split}
\ee
where the shorthand $z^{ij..} = z^i z^j...$ is used. Thus, with $\a>0$ and $\b^2\ge\a$ we are guaranteed a positive definite metric. To study the bounds of $\b^2$ in more detail, we must analyse the eigenvalues of $G$, which is computationally expensive in this example. 

$\d K$ is of the form \eqref{eq:dKgen}, and thus contributes to the first, third and fourth torsion classes. In addition, we find that $\gamma$ does not set the phase of $\d \Omega$, which shows that $\cW_5$ is non-zero.  All in all, the torsion classes are given by \eqref{sctvtorsions}, where in the patch $z_1, z_4, z_6 \neq 0$  
\bea
\nn
&\cW_1^0 = -i \frac{2 B_1B_2\sqrt{\mbox{det} g_{ab}}}{3 \a \b p}\\ \nn
&\cW_2^0= \cW_1^0 \left\{
(2\b^2 - \a)\left(
J +\frac{3i \beta^2}{2} K \w \bar{K}
 \right) + \a^2 \tilde{\xi}^3 \left(-3 \frac{|K_1|^2|K_2|^2}{\mbox{det}g_{ab}} j + \frac{i}{|z^6|^2} \cD z^5 \w \cD \bar{z}^5 \right)
\right\}
 \\
&\cW_3^0 = -\cW_4^0   \wedge \left(J + i \b^2 K \wedge \bar{K} \right) 
\\ \nn
&\cW_4^0 =  \frac{1}{4\b^2} \left\{\d \ln p + \frac{2 (B_1^2 \tilde{\xi}^2 -B_2^2 \tilde{\xi}^1)}{p}  \mathrm{Re}(\bar{z}_5 \mathcal{D}z^5)\right\} \\ \nn
&\cW_5^0 = 2 \b^2 \cW_4^0 + \d \ln p - \frac{1}{2} \d \ln \mbox{det} g_{ab} \; .
\eea
For any constant $B_1, B_2$, $\cW_4^0$ and $\cW_5^0$ are closed, and hence exact since $b^1=0$. Primitivity of $\cW_2^0$ and $\cW_3^0$ is readily checked.

Since both $\cW_4^0$ and $\cW_5^0$ are exact, we can choose $\a$, $\b$ to put a linear combination of them to zero. However, since the Lie forms are not proportional, there is no parametric choice that gives a half-flat $SU(3)$ structure. Similarly, complex and symplectic $SU(3)$ structures cannot be reached. Comparing with the maximally symmetric string vacua of section \ref{sec:survey}, we thus see that this $SU(3)$ structure does not allow $\cN=1$ vacua. Calibrated $\cN=0$ vacua, however, may be allowed. Particularly, by choosing  $\a, \b$ and $\g$, we can set $\d \mbox{Re}(e^{-2 \phi} \Omega) = 0 = \d (e^{-2\phi} J \w J)$ as required for  calibrated vacua of type IIB (with O5 planes) or heterotic string theory \cite{Lust:2008zd,Held:2010az}. These constraints can be met without violating the positivity of the $SU(3)$ metric. Moreover, although $\cW_2^0$ is slightly more complicated than for $\mathbb{CP}^3$, its form is similar to that required for calibrated vacua of no-scale type listed in table \ref{tab:su3N0}.  We leave a more detailed investigation of this question, as well as the other constraints for calibrated vacua, to the future.

\subsubsection{Additional examples of $\mathbb{CP}^1$ bundles}

Finally, let us present some data on $SU(3)$ structure on $\mathbb{CP}^1$ bundles over two-dimensional SCTVs with three, five and six generators. For each example, we present a choice of $K$ that meets the three conditions specified in section \ref{sec:su3constr}, thus confirming that such a form can be found on toric $\mathbb{CP}^1$ bundles. In contrast to the previous examples, we have not been able to find $K$ whose exterior derivatives of the are of the form \eqref{eq:dKgen}, nor lead to $SU(3)$ structure with exact  the Lie forms. In all other respects, the analysis parallels the previous section, so we will only present the results of our study. 

\paragraph{$\mathbb{CP}^1$ over $\mathbb{CP}^2$:}

When viewed as a twistor space or a coset, $\mathbb{CP}^1$ over $\mathbb{CP}^2$ allows a half-flat $SU(3)$ structure \cite{Tomasiello:2007eq,Koerber:2008rx}. Consequently, one would expect there to be an equally simple choice for $K$ as there is on $\mathbb{CP}^3$. Curiously, such a simple $K$ has not been found.  The reason is that none of the $P_{ij}$ eigenvectors presented in table \ref{tab:1} have nowhere vanishing norm, and so a rather involved linear combination of $K_i$ is needed to construct $K$.  

One possibility is
\be
\hat{K} =  \alpha_2 \sqrt{\xi_2} K_2 + \alpha_3 K_3 \; .
\ee
If $\alpha_2$, $\alpha_3$ are pure phases then the norm of  this form is $|\hat{K}|^2 = \mbox{det}g_{ab} \neq 0$. We note that the exact contributions that the non-constant $\mbox{det}g_{ab}$ give to the Lie forms can be compensated by choices of $\a$ and $\b$ as in \eqref{eq:dpcontr}.

To satisfy also the charge requirement on $K$, one possible choice is to take
\be \label{eq:aphase}
\alpha_2 = z^4/|z^4|  \; \; , \; \; \alpha_3 = |z^4|/z^4 \; .
\ee
This choice of $K$ is valid for any odd value of the twist parameter $n^1$. Computing the torsion classes for undetermined $\alpha_i$ is a daunting task, as $\d K$ is not of the simple form \eqref{eq:dKgen}. Specialising to $n^1=1$ and using \eqref{eq:aphase}, we find they are all non-vanishing, and $\cW_{4,5}$ are not closed. The expressions  for the torsion classes are not particularly illuminating, so we do not reproduce them here.

\paragraph{Toric $\mathbb{CP}^1$ bundle, $N=5$: }

This example is quite similar to the $\mathbb{CP}^1$ bundle over $\mathbb{F}_0$. $K$ can be constructed using the orthogonal forms $K_1$ and 
\be \label{eq:k2perp}
K_2^{\perp} = K_2-\frac{\bar{K_1}\cdot K_2}{|K_1|^2} K_1 \; ,
\ee
where $K_{1,2}$ can be found in table \ref{tab:1}. $\widetilde{\Omega}$ has even charge if and only if $n^1, n^3+a$ are odd and $n^2$ is even. The (1,0)-form 
\be
\hat{K}= z^7 K_1 + z^6 K_2^{\perp}
\ee
then has the right charge $Q (z^7 K_1) = Q (z^6 K_2) = \frac{1}{2} Q (\widetilde{\Omega})$ if we impose $n^1 = -3$, $n^2= 2$, and $n^3=a-1$.  With these parameter values, we identify the
basis of the Mori cone and the corresponding charge basis $\widetilde{Q}^1=Q^1-Q^3$, $\widetilde{Q}^2=Q^2-n^2Q^4$, $\widetilde{Q}^{3,4}=Q^{3,4}$ (see appendix \ref{sec:Mori}), and use the result to show that the norms $|K_{1,2}|$ are non-zero whenever $a\le-1$. Since $z^6$ and $z^7$ cannot be zero simultaneously, we have then constructed a $K$ that has the required properties.

\paragraph{Toric $\mathbb{CP}^1$ bundle, $N=6$: }

There are three two-dimensional SCTVs with six generators, and hence we get three different three-dimensional $\mathbb{CP}^1$ fibrations. Here we show how $K$ can be chosen for the first of these. The analysis for the two other examples is completely analogous.

We construct $K$ using $K_{1,2}$ from table \ref{tab:1}. Since these are not orthogonal, we first define $K_2^{\perp}$ as in \eqref{eq:k2perp}. Inspired by the previous examples, we take
\be
\hat K = z^7 K_1 + z^8 K_2^{\perp}~.
\ee 
and impose the charge constraint $Q (z^7 K_1) = Q (z^8 K_2) = \frac{1}{2} Q (\widetilde{\Omega})$. This equation has the solution
\be \label{eq:6charge}
n^1 = -1; \quad n^2= a-2; \quad n^3=2; \quad n^4 = a-1
~.
\ee
To show that the norm of $\hat K$ is non-vanishing, we note that $z^7$ and $z^8$ cannot be zero simultaneously. Moreover, the norms of $K_1$ and $K_2^{\perp}$ are bounded from below by the K\"ahler moduli $\widetilde{\xi}^a$, where the $U(1)$ charge basis associated with the parameter values \eqref{eq:6charge} is $\widetilde{Q}^2=Q^2-Q^4$, $\widetilde{Q}^3=Q^3 - n^3 Q^5$, $\widetilde{Q}^{1,4,5}=Q^{1,4,5}$.

\section{Discussion}
\label{sec:conclusion}

Six-dimensional $SU(3)$ structure manifolds have a long history in string theory compactifications, and have been used to construct a variety of four-dimensional vacua where supersymmetry is either preserved or spontaneously broken. Since $SU(3)$ structure manifolds can accommodate fluxes, these vacua are believed to have fewer moduli than vacua arising from compactifications on CY manifolds. However, confirming this assertion is difficult, since in contrast to the great number of CY manifolds, comparably few explicit examples of $SU(3)$ structure manifolds exist. One obstacle in the construction of example manifolds is the lack of integrable complex structures which hinders the use of algebraic geometry. In this paper, we have used the fact that toric varieties allow both integrable and non-integrable almost complex structures to construct new examples of $SU(3)$ structure manifolds. In doing so we show that  the construction of \cite{Larfors:2010wb} extends to an infinite class of toric varieties, which is an important step to a more systematic study of toric $SU(3)$ structures.

We have shown that $\mathbb{CP}^3$ and all toric $\mathbb{CP}^1$ fibrations allow $SU(3)$ structures, since they have even first Chern class.\footnote{For toric $\mathbb{CP}^1$ fibrations, this is true if the parameters of the associated fan are chosen accordingly.} In contrast, toric $\mathbb{CP}^2$ fibrations do not allow $SU(3)$ structures. The $SU(3)$ structures can be constructed using the method of \cite{Larfors:2010wb}, which is based on a local one-form $K$, and we argue that this form exists as long as $c_1$ is even. Indeed, we have constructed $K$ explicitly for $\mathbb{CP}^3$, and $\mathbb{CP}^1$ bundles over two-dimensional SCTVs with up to six generators. These $K$ are not claimed to be unique, and in two of the examples we investigate how simple modifications of $K$ lead to changes in the torsion classes.  In general, we have found that the torsion classes simplify if $d K$ satisfies the relation \eqref{eq:dKgen}. A better understanding of the relation between the choice of $K$ and the resulting torsion classes would certainly be desirable, and we hope to return to this in the future. It would also be interesting to investigate if alternative methods of constructing $SU(3)$ structures can be used to derive global constraints on the torsion classes. In this respect, it is interesting to note that $\mathbb{CP}^1$ fibrations over four-dimensional Riemannian spaces are twistor spaces, so it is possible twistor techniques can be used in such studies.

\begin{figure}
\centering
\includegraphics[width=9cm]{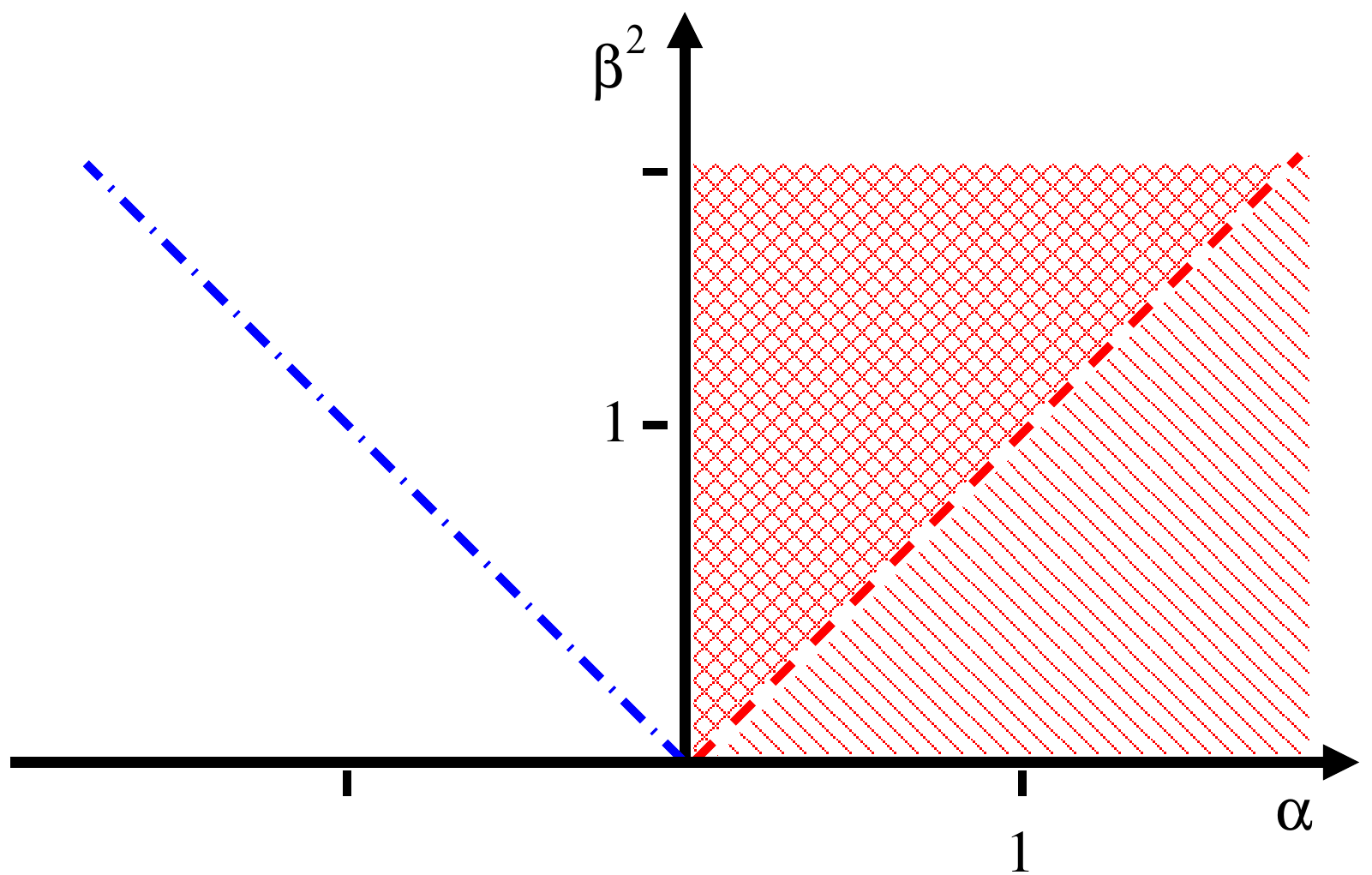}
\caption{\it Toric $SU(3)$ structures are parameterised by the real functions $\alpha$ and $\beta$. Metric positivity always restricts the parameters to the shaded area $\alpha>0$, and sometimes further to $\beta^2\ge\alpha$. The dash-dotted blue line is the symplectic limit $J = \alpha \tilde{J}|$, which is clearly excluded by metric positivity.}
\label{fig:parbound}
\end{figure}

Since the method we use has a parametric freedom, specified by three real functions $\a, \b$ and $\g$, it is possible to tune toric $SU(3)$ structures to some degree. In accordance with \cite{Larfors:2010wb}, we  show that the exterior derivative of $J$ is proportional to $(\a+\b^2)$ when $\a$ and $\b$ are constant. Moreover, we find that the phase of $\cW_1$ and $\cW_2$ is set by $\g$ and that exact contributions to the Lie forms $\cW_4$ and $\cW_5$ can to some extent be compensated by $\a$ and $\b$; one exact Lie form can always be set to zero by a judicious choice of parameters, and if in addition the quotient $\cW_4/\cW_5$ is constant, both $\cW_4$ and $\cW_5$ can be removed.\footnote{Since SCTVs have vanishing odd Betti numbers, it is enough to prove that the Lie forms are closed to ascertain that they are exact.} However, an important constraint on the parametric freedom of  $SU(3)$ structures comes from positivity of the associated metric. We have shown that the $SU(3)$ metric is related to the metric inherited from $\mathbb{C}^n$ by
\be  \nn
G_{mn} =\alpha \left[ \widetilde{G}_{mn} + \left(\frac{\beta^2}{\alpha}-1\right) \mbox{Re} \left(K_m \bar{K}_{n}
\right) \right]
\; ,
\ee
so that metric positivity requires $\a > 0$. The parameter limit $\a = - \b^2 < 0$ is thus not attainable, and the toric $SU(3)$ structures do not have a generic symplectic limit, contrary to what the expression for the torsion classes would suggest (as shown in figure \ref{fig:parbound}). 

To complement our general analysis,  we compute the torsion classes in full for three examples, and find that all are in general non-vanishing. The toric $SU(3)$ structure we construct on $\mathbb{CP}^1 \hookrightarrow \mathbb{CP}^2$ has non-exact Lie forms and so does not agree with the half-flat $SU(3)$ structure found in previous studies. Contrarily, on $\mathbb{CP}^3$  and $\mathbb{CP}^1 \hookrightarrow \mathbb{F}_0$, we show that $K$ can be chosen so that the Lie forms are exact, that $|\cW_2| \propto |\cW_1|$, and that, for constant $\a, \b$, $\cW_3 = -\cW_4 \w (J + i \b^2 K \w \overline{K})$.  On $\mathbb{CP}^3$, $K$ can be simplified further, leading to a restricted half-flat $SU(3)$ structure, in accordance with previous studies. The $SU(3)$ structure on $\mathbb{CP}^1 \hookrightarrow \mathbb{F}_0$ is less adaptable, and always retain non-zero $\cW_1$ and $\cW_2$, in addition to at least one of the Lie forms. 
 
The existence of toric $SU(3)$ structures opens up for many applications, even though contrary to the CY case, an $SU(3)$ structure is not enough to prove that string compactification results in a four-dimensional vacuum. In many cases, the equations that define string vacua can be translated to necessary constraints on the torsion classes of the $SU(3)$ structure, and our example manifolds can be compared with these constraints.  In particular, it is a well-known fact that the restricted half-flat $SU(3)$ structures on $\mathbb{CP}^1 \hookrightarrow \mathbb{CP}^2$ and $\mathbb{CP}^3$ matches the requirements for several vacua, including supersymmetric ones. For $\mathbb{CP}^3$, the less constrained choices for $K$ mentioned above do not lead different types of string vacua.

We have not found any new SCTVs that match the conditions for supersymmetric string vacua. However, we have found that $\mathbb{CP}^1 \hookrightarrow \mathbb{F}_0$ matches at least some of the necessary constraints for calibrated $\cN=0$ vacua, if the parameters $\a, \b, \g$ are chosen accordingly. A more complete study is needed to see if all Bianchi identities and equations of motion for such vacua are satisfied, and we hope to come back to this in the future. It would be also be interesting to investigate if other non-supersymmetric string vacua can be constructed on this manifold. Of particular interest are dS vacua, which are notoriously difficult to find in string theory. Such vacua require negative scalar curvature of the internal space, so it is interesting to note that $\a, \b$ can be chosen so that the contribution from the Lie forms to the scalar curvature of $\mathbb{CP}^1 \hookrightarrow \mathbb{F}_0$ is negative definite.  On the other hand, this $SU(3)$ structure is neither half-flat nor symplectic, as has been assumed for known dS solutions. Consequently, a new take on such constructions would be required to investigate whether this toric $SU(3)$ structure could be of relevance for dS vacua in string theory.

\subsection*{Acknowledgements}

This research is supported by the Swedish Research Council (VR) under the contract 623-2011-7205. It's a pleasure to thank  R.~Bryant, P.~Candelas, X.~de~la~Ossa, R.~Davies, E.~Sharpe and D.~Tsimpis for illuminating discussions at various stages of this project. Additionally, I'm grateful to D.~Andriot, J.~Bl{\aa}b\"ack, U.~Danielsson and G.~Dibitetto for interesting remarks regarding dS vacua on $SU(3)$ structure manifolds. 

\newpage
\begin{appendix}

\section{Mori and K\"ahler cones}\label{sec:Mori}

For the constructions of $K$, it is crucial to know that the K\"ahler moduli $\xi^a$ are strictly positive. This condition is satisfied within the K\"ahler cone for any non-singular manifold. However, the identification of K\"ahler moduli depends on the choice of $U(1)$ charges $Q^a$, as we now explain. To do so, we need to introduce some concepts from complex geometry. Most of the material in this appendix, which is included for the reader's convenience, can be found in the pedagogical review \cite{Denef:2008wq}.

A divisor $D$ is a formal sum of holomorphic hypersurfaces $S^I$, that are defined locally in a coordinate patch $U^{\alpha}$ by a holomorphic equation $f^I_{\alpha}=0$, such that $f^I_{\alpha}/f^I_{\beta}$ has no zeros or poles on the intersection $U^{\alpha} \cap U^{\beta}$. For a toric variety, each $z^i$ defines a divisor:
\be
D_i: z^i=0,~~~~ i=1,\dots 6~.
\ee
Divisors are linearly equivalent, $D_i = D_j$, if the quotient of their defining equations is a globally defined rational function. Using this, the $U(1)$ gauge invariances can be used to show that there are only $s$ linearly independent divisors $D_i$ on an SCTV, where $s$ is the number of $U(1)$ actions.

Transversal intersections of divisors, $D_i D_j$, are holomorphic curves. The integral of the K\"ahler form over a holomorphic curve measures the area of the curve, and is therefore positive:
\be
\int_C J \ge 0,~~~~ C\mbox{ holomorphic curve}.
\ee
It can be shown that the intersections $D_i D_j$ generate the full set of two-cycle classes with holomorphic representatives, which is known as the Mori cone. However, since not all divisors need be linearly independent, not all $D_i D_j$ are linearly independent either. One can define a basis $C^a$, $a=1,2,..s$, so that all $D_i D_j$ can be expanded in $C^a$ with non-negative coefficients:
\be \label{eq:didj}
D_i D_j = \sum_{a=1}^s b_{ij}^a C^a
\quad \mbox{where} \quad
b_{ij}^a \ge 0 
~.
\ee
The $C^a$ constitute a basis for the Mori cone, and we can use them to define K\"ahler moduli
\be \label{eq:kahmod}
\xi^a = \int_{C^a} J \ge 0~.
\ee
Here we have used that the $C^a$'s are holomorphic curves to infer that $\xi^a$ are positive.

Given a basis $C^a$ for the Mori cone, one can always change the charge basis so that 
\be \label{eq:Ca}
D_i C^a=Q^a_i
~.
\ee
It is important to note that it is only when we express the moment maps in terms of this charge basis, that we can conclude that the parameters $\xi^a$ match \eqref{eq:kahmod}. If there are parameters in the charges $Q^a$ ($a, n^a$ in the examples above), their sign must be determined before it can be concluded that the expansion coefficients $b_{ij}^a$ in \eqref{eq:didj} are strictly positive. In other words, which charge basis is associated with $C^a$ depends on whether the parameters $a, n^a$ are positive or negative.

\end{appendix}
%


\newpage
\bibliographystyle{JHEP}

\providecommand{\href}[2]{#2}\begingroup\raggedright\endgroup

\end{document}